\documentclass[fleqn,usenatbib]{mnras}

\usepackage{newtxtext,newtxmath}

\usepackage[T1]{fontenc}
\usepackage{float}
\usepackage{booktabs}
\usepackage{natbib}
\usepackage{subcaption} 
\usepackage{longtable}
\DeclareRobustCommand{\VAN}[3]{#2}
\let\VANthebibliography\thebibliography
\def\thebibliography{\DeclareRobustCommand{\VAN}[3]{##3}\VANthebibliography}

\usepackage{graphicx}	
\usepackage{amsmath}	

\title[Fullerenes in the shell of cLBV WRAY~16-232]{Discovery of Fullerenes in the shell of candidate Luminous Blue Variable WRAY~16-232 }
\author[R. Arun et al.]{
R. Arun,$^{1}$\thanks{E-mail: arunroyon@gmail.com}
S. A. Prasoon$^{1,2}$,
Blesson Mathew$^{2}$,
D. Akhila$^{2,4}$,
Gourav Banerjee$^{1}$,
B. Shridharan$^{3}$,
\newauthor
G. Maheswar$^{1}$ and
Arun Surya$^{1}$
\\
$^{1}$Indian Institute of Astrophysics, Sarjapur Road, Koramangala, Bangalore 560034, India\\
$^{2}$Center of Excellence in Astronomy and Astrophysics, Department of Physics and Electronics, CHRIST (Deemed to be University), Bangalore 560029, India\\
$^{3}$Tata Institute of Fundamental Research, Homi Bhabha Road, Mumbai 400005, India\\
$^{4}$Instituto de Astronomía y Ciencias Planetarias, Universidad de Atacama, Copayapu 485, Copiapó, Chile
}

\date{Accepted XXX. Received YYY; in original form ZZZ}

\pubyear{\the\year{}}

\begin{document}
\label{firstpage}
\pagerange{\pageref{firstpage}--\pageref{lastpage}}
\maketitle

\begin{abstract}
We report the discovery of fullerene in the circumstellar environment of WRAY~16-232, a strong candidate luminous blue variable. Multiple pointings of archival \textit{Spitzer} IRS spectra reveal, for the first time, the presence of prominent vibrational bands of C$_{60}$ at 17.4 and 18.9~$\mu$m in an LBV envelope, along with the strong polycyclic aromatic hydrocarbon features. These observations suggest that, despite the harsh radiative conditions, large carbonaceous molecules can form, process and survive in the ejecta of massive stars. Complementary optical spectroscopy with SALT HRS shows multiple P\,Cygni profiles in H$\alpha$, He\,\textsc{i}, and Fe\,\textsc{ii} lines, which are indicative of a dense, expanding wind and substantial mass loss. Furthermore, analysis of decade long photometric data shows short-term brightness variations of $\sim$ 0.5~mag. These results not only reinforce the classification of WRAY~16-232 as a strong LBV candidate but also provide new insights into the mechanisms of dust formation and the chemical enrichment of the interstellar medium by massive stars. We discuss various scenarios for fullerene formation in such environments, and find that shock processing due to wind-wind interactions could be playing a vital role.The shell of WRAY~16-232 has an ideal UV field strength and the time scales appears to match with shock processing timescales. The results highlight the need for further high spatial/spectral resolution and temporal observations to confirm the formation and survival scenario of C$_{60}$ in its shell.
\end{abstract}




\section{Introduction}
\label{intro}

Luminous Blue Variables (LBVs) represent a transient yet pivotal stage in the evolution of massive stars. Exhibiting dramatic photometric and spectroscopic variability \citep{1994PASP..106.1025H}, LBVs serve as a bridge between main-sequence OB stars and later evolutionary stages such as Wolf-Rayet (WR) stars or other supernova progenitors. For stars with initial masses in the range $25M_{\odot} < M < 40M_{\odot}$, the LBV phase occurs post-Red Supergiant (RSG) evolution \citep{refId0}, while for masses between $40M_{\odot} < M < 60M_{\odot}$, LBVs act as a precursor to the WR phase \citep{2014A&A...564A..30G}. Their powerful stellar winds expel CNO-processed material into the interstellar medium, thereby influencing star formation and galactic chemical enrichment \citep{VINK2008419,Joachim2008A&ARv..16..209P}. The brief lifespan of the LBV phase ($\sim 10^{4}$--$10^{5}$ years; \citealp{1991IAUS..143..485H}) contributes to their rarity, making statistical analyses challenging. Given the small sample size of LBVs \citep{Richardson2018RNAAS...2..121R}, even a few misclassifications can significantly hinder our understanding of the role these objects play in massive star evolution, emphasizing the importance of applying rigorous and consistent observational criteria \citep{2014ApJ...790...48H}.

While LBVs are known for their complex circumstellar environments, the presence of fullerenes in such settings has never been reported. Fullerenes are a unique family of stable, cage-like molecules made entirely of carbon atoms, usually in even numbers. One of the most well-known and stable members of this family is buckminsterfullerene (C$_{60}$), commonly referred to as buckyballs. This remarkable molecule is composed of 60 carbon atoms arranged in a spherical shape, featuring 12 pentagonal and 20 hexagonal faces. C$_{60}$ exhibits icosahedral symmetry, the highest point group symmetry for a three-dimensional molecule, making it an extremely stable and chemically resilient molecule \citep{doi:10.1021/ja00180a055}. 

Fullerenes were first identified in the laboratory by \citet{1985Natur.318..162K}. The first confirmed astronomical detection of C$_{60}$ was made by \citet{doi:10.1126/science.1192035} in the planetary nebula (PN) Tc 1.  Since then, several studies have detected fullerenes under a range of astrophysical conditions. C$_{60}$ has been detected in other PN \citep{García-Hernández_2010,García-Hernández_2012,10.1093/mnras/stt2070}, in pre-main-sequence stars such as Herbig Ae/Be stars \citep{10.1111/j.1365-2966.2012.20552.x,10.1093/mnras/stad1511}, in reflection nebulae (RNe) like NGC 7023 and NGC 2023 \citep{Sellgren_2010}, in the hydrogen-deficient and carbon-rich environments of R CrB stars \citep{2011ApJ...729..126G}, and in post-AGB stars \citep{10.1093/mnras/stx1237}.

C$_{60}$ formation is typically associated with the interplay of high-temperature environments and carbon-rich chemistry, but its survival in the harsh radiation fields of massive stars remains an open question \citep{Berne2012PNAS..109..401B}.
There are two primary formation mechanisms for C$_{60}$ in the literature. The first is the "top-down" process, in which large Polycyclic Aromatic Hydrocarbons (PAH) molecules (with $N_c \geq 60$) lose their hydrogen due to UV irradiation, leading to the formation of graphenes. Continued UV exposure causes graphenes to shed C$_2$ units, eventually folding into closed cage structures and stabilizing as C$_{60}$ due to their exceptional molecular stability \citep{2006JChPh.125d4702I,2015A&A...577A.133B}. Also, the "top-down" process can also occur when photochemical processing of Hydrogenated Amorphous Carbon (HAC) grains, resulting in the formation of PAHs and fullerenes (\citealp{García-Hernández_2010,Micelotta_2012,Bernard-Salas_2012}). The second mechanism is the “bottom-up” process, in which small carbon-chain molecules and PAHs, subjected to UV irradiation at moderate temperatures in carbon-rich and hydrogen-poor environments \citep{Crichton2022}, assemble into fullerenes and larger PAHs, akin to laboratory synthesis methods. Beyond these linear routes, recent studies indicate a more dynamic chemical network, in which PAHs undergo fragmentation into smaller clusters (C$_n$, C$_n$H$_m$), which then isomerize or coalesce into cage-like structures before stabilizing as fullerenes \citep{Omont2021}.

While several studies have reported the detection of PAHs in LBVs and WR stars \citep{1999A&A...341L..67V, Rajagopal_2007, Jiménez-Esteban_2010, 10.1093/mnras/stx563, Rizzo_2023}, investigations of fullerenes have predominantly focused on evolved low-mass stars \citep{García-Hernández_2012}, with limited evidence for their presence in massive stars such as LBVs. To date, C$_{60}$ has only been detected in regions where physical conditions vary over much longer timescales, in contrast to the dynamic environments of massive stars like LBVs. These stars experience significant mass loss and strong stellar winds during their S Doradus (S Dor) cycles, which typically span 5–10 years and can extend beyond 20 years \citep{2001A&A...366..508V}. The detection of C$_{60}$ in the harsh and highly variable LBV environment might hint towards a potentially different formation mechanism, where C$_{60}$ is synthesized and sustained even in the extreme conditions of LBVs. 

In this study, we report the detection of C$_{60}$ emission in Spitzer Infrared Spectrograph (IRS) spectra of WRAY~16-232, a previously identified candidate LBV (cLBV). Located in the constellation Ara at a distance of 2.1 $\pm$ 0.5 kpc \citep{2023A&A...674A...1G}, WRAY~16-232 was first classified as a cLBV by \citet{Gvaramadze10.1111/j.1365-2966.2010.16496.x} based on archival 24 $\mu$m data from the Spitzer/MIPSGAL survey \citep{Carey_2009}. It is designated [GKF2010] MN46, consistent with other mid-infrared (MIR) envelope sources discovered using Spitzer. Prior to this classification, the object had been assigned various types in the literature.

To our knowledge, WRAY~16-232 first appeared in the literature by \citet{1966PhDT.........3W}, where it was initially classified as a likely PN. However, \citet{1966PASP...78..136W} later rejected this classification, and \citet{1967ApJS...14..125H} labeled it as a doubtful PN. Over time, the classification of WRAY~16-232 has changed considerably. \citet{1973ApJ...185..899S} characterized the source as a "very steep Balmer decrement object." Subsequently, \citet{1982MNRAS.199.1017A} classified it as a highly obscured Be star, while \citet{1983RMxAA...8...39M} identified it as an M-type supergiant. \citet{1992secg.book.....A} later described it as a peculiar emission-line star.
Recently, an interferometric study by \citet{2022A&A...657A...4M} reported the presence of a companion star to WRAY~16-232 at an angular separation of 0.3 mas, corresponding to a linear separation of approximately 0.63 au \citep{2023A&A...674A...1G}. The companion is 3.7 mag fainter than the primary star and is well within the circumstellar nebula. The circumstellar nebulae around WRAY~16-232 identified in Spitzer 24 $\mu$m image appears to be spherical and has a size of 0.49 x 0.49 pc \citep{Gvaramadze10.1111/j.1365-2966.2010.16496.x, 2022A&A...657A...4M}.

This paper is organized as follows: Section~\ref{sec:data} describes the observational data, Section~\ref{sec:analysis} outlines our analysis, and Sections~\ref{sec:discussion} and~\ref{sec:conclusion} provide discussions and conclusions, respectively.

\section{Data}\label{sec:data}

While compiling the Spitzer IRS spectra of LBVs and cLBVs listed in \cite{Smith2019} and \cite{2022A&A...657A...4M}, we identified WRAY~16-232 to have three distinct observations, covering the central star along with the southern and northern part of the star's envelope. This spatial configuration offers a unique opportunity to investigate the dust formation processes within the circumstellar shell of this LBV candidate. The following subsections describe the specific datasets used to study WRAY~16-232 and its surrounding envelope in detail.

\subsection{Spitzer Observations}

\subsubsection{Imaging}
The MIR images of WRAY~16-232 were examined using archival Spitzer data obtained from the Infrared Processing and Analysis Center (IPAC\footnote{https://www.ipac.caltech.edu/doi/irsa/10.26131/IRSA433. IPAC is hosted by California Institute of Technology (Caltech) and operates under contract with National Aeronautics and Space Administration (NASA).}). We made use of data from the Infrared Array Camera (IRAC; \citealp{Fazio2004ApJS..154...10F}) and the Multiband Imaging Photometer for Spitzer (MIPS; \citealp{2004ApJS..154...25R}). The Spitzer IRAC 4 image at 8.0 \micron\ were obtained from  Galactic Legacy Infrared Mid-Plane Survey Extraordinaire (GLIMPSE; \citealp{Benjamin2003PASP..115..953B}) survey and the MIPS1 image at 24 \micron\ were obtained from Multiband Imaging Photometer for Spitzer Galactic Plane Survey (MIPSGAL: \citealp{Carey_2009}). The IRAC 4 data has higher-resolution ($\sim$ 2\arcsec) than MIPS 1 ($\sim$ 6\arcsec) data. All Spitzer data were retrieved as pipeline-processed mosaic products.

\subsubsection{Spectroscopy}
The MIR spectra of WRAY~16-232 were retrieved from the Combined Atlas of Sources with Spitzer IRS Spectra (CASSIS\footnote{CASSIS is a product of the IRS instrument team, supported by NASA and JPL.}) archive \citep{2011ApJS..196....8L,2015ApJS..218...21L}. The spectra were obtained as part of the Spitzer program titled `IRS Investigation of 24 Micron Compact Ring Sources' (Program ID: 50808, PI: Sean Carey). Observations were carried out on 2009-04-21 at 04:47:53 UTC and are identified by the Astronomical Observation Request (AOR) Key 32555008.

The dataset includes three distinct pointings, separated by 0.79$^{\prime\prime}$ (center), 44.13$^{\prime\prime}$ (north), and 47.34$^{\prime\prime}$ (south) from the central target. The spectra were acquired using the low-resolution (LR) mode of the IRS instrument, covering the wavelength range of 7.8–35~$\mu$m. The 6–7.5~$\mu$m region was not observed in this program.

The raw spectra were processed using the enhanced offline pipeline CASSISjuice \citep{lebouteiller2023cassisjuiceopensourcepipelineoffline}. CASSISjuice improves data usability by organizing pointings under a unified target ID and incorporating a robust and consistent pipeline for all IRS staring-mode observations. 

\subsection{SALT HRS Spectroscopy} \label{sec: SALT_spectro}
The optical spectra analyzed in this study were obtained from the South African Large Telescope (SALT) High-Resolution Spectrograph (HRS) \citep{2008SPIE.7014E..0KB,2010SPIE.7735E..4FB,2012SPIE.8446E..0AB,2014SPIE.9147E..6TC} archive. The data were taken on 2016-05-16 using the HRS in its low-resolution mode, which provides a resolving power of \( R \sim 16,000 \) with a exposure time of 2400 sec (Prop ID: 2016-1-SCI-012 - PI:Alexei Kniazev). It is important to note that seven spectra of WRAY~16-232, spanning a period of seven years (May 2016 to September 2022), are available in the SALT archive. For the detailed line analysis, we use the observation closest in time to the \textit{Spitzer} observation. Since the primary aim of this article is to report the discovery of fullerenes in WRAY~16-232, we did not undertake a comprehensive multi-epoch analysis. Nevertheless, we performed a preliminary investigation of the H$\alpha$ line to assess the variable nature of WRAY~16-232.

The reduced data of WRAY~16-232 is taken from the SALT archive\footnote{https://ssda.saao.ac.za/}. The primary reduction such as overscan correction, bias subtractions and gain correction, was carried out with the SALT science pipeline \citep{2010SPIE.7737E..25C}. And the spectroscopic reduction of the HRS spectra is reduced using the MIDAS pipeline \citep{Kniazev_2016,Kniazev2017ASPC..510..480K}, which includes standard procedures such as bias subtraction, flat-fielding, wavelength calibration, and cosmic ray removal. These archival reductions ensured that the spectra were ready for scientific analysis, removing the need for additional data processing. Due to the very low signal-to-noise ratio in the blue arm (3900--5500~\AA), these data were excluded from the analysis. Our study is therefore based on the red arm spectra, covering the 5400--8900~\AA\ range.

\subsection{Photometry}

To study the photometric variability, we compiled the available magnitudes for WRAY~16-232.The photometric magnitudes, along with stellar parameters of the star is mentioned in Table \ref{tab:photo}. There are significant difference between the B and R band magnitudes, with the B-band being 4.2 times dimmer than the R-band, indicating high extinction towards the star.

\begin{figure*}
    \includegraphics[width=\columnwidth]{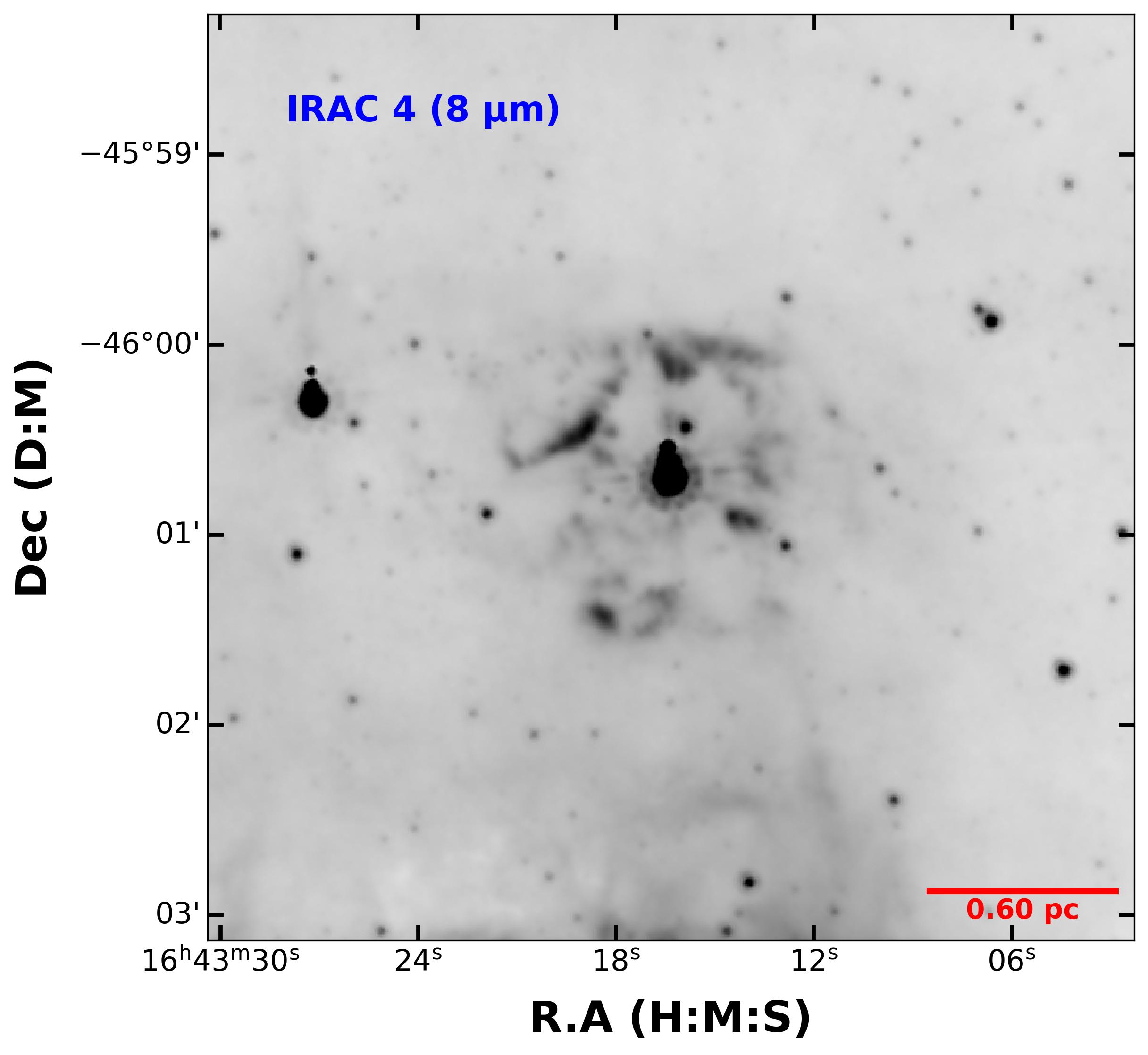}
    \includegraphics[width=\columnwidth]{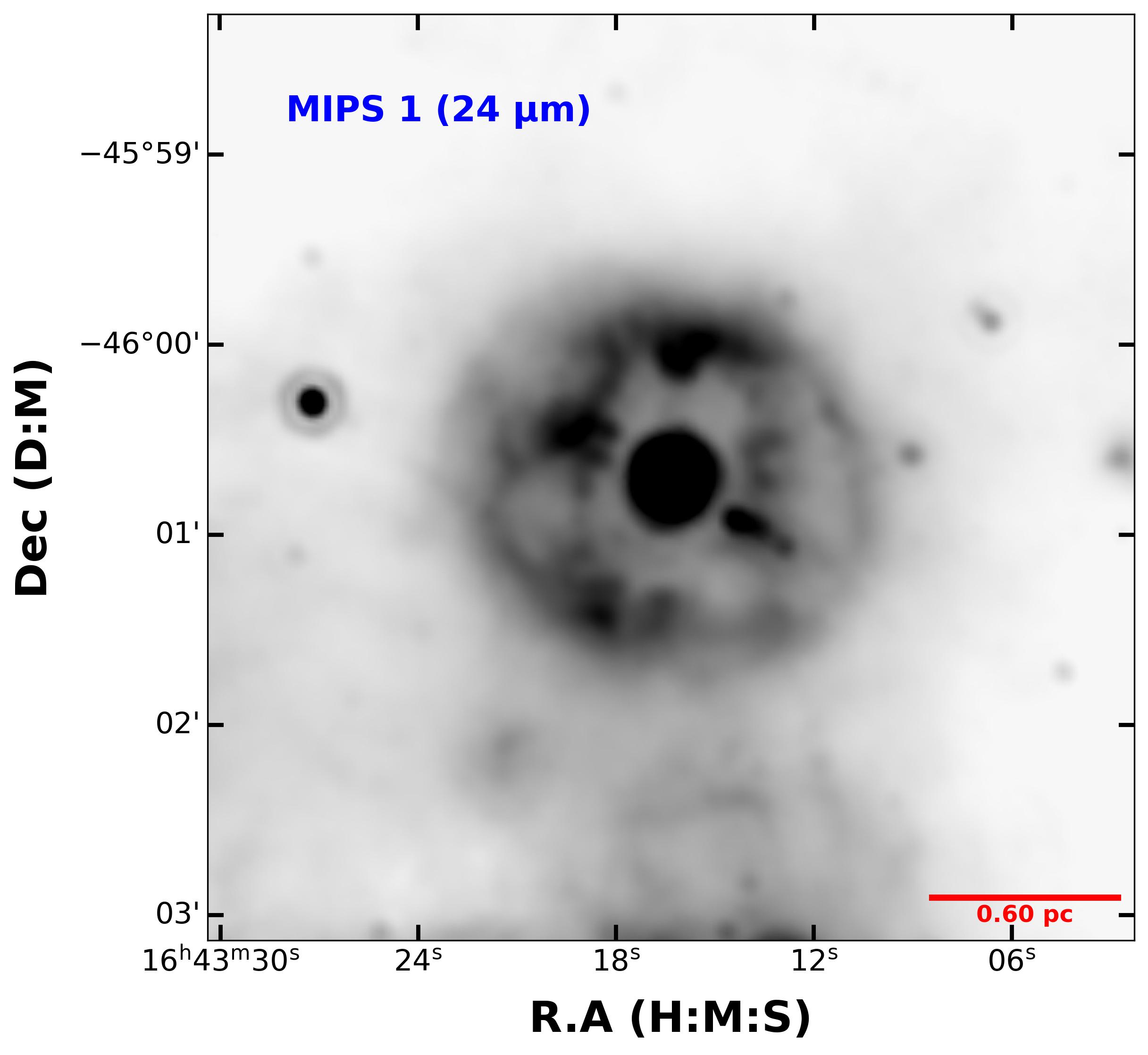}
    \caption{
    Infrared images of WRAY~16-232 showcasing its circumstellar environment. \textbf{Left:} Spitzer IRAC 4 (8 $\mu$m) image highlighting the distribution of PAHs and warm dust in the nebula. \textbf{Right:} Spitzer MIPS 1 24 $\mu$m image emphasizing thermal emission from cooler dust grains in the circumstellar shell. Both images are oriented with North up and West to the left. The nebula exhibits a nearly spherical morphology, traced by the 24 $\mu$m cool dust, with clumpy structures and arc-like features more pronounced in 8 $\mu$m, suggesting complex mass-loss processes and wind interactions.}
    \label{fig:irs_image}

\end{figure*}

To investigate the photometric variability of WRAY~16-232, we retrieved \textit{Gaia} G-band, BP-band and RP-band photometric data \citep{2016A&A...595A...1G,2023A&A...674A...1G}, spanning from October 2014 to March 2017. Additionally, V-band photometric data were obtained from the All-Sky Automated Survey for Supernovae (ASAS-SN; \citealp{Shappee_2014,Jayasinghe_2019}), observed between March 2016 and October 2018 \citep{Kochanek_2017}. Notably, a brightening event is evident in both the \textit{Gaia} G-band and \textit{ASAS-SN} V-band light curves between January 2017 and March 2017. We also incorporate \textit{r}-band and \textit{i}-band photometry from the Bochum Galactic Disk Survey (GDS), as presented in \citet{Hackstein2015AN....336..590H}. It provides time-series observations of the Southern Galactic plane. The \textit{r}- and \textit{i}-band magnitudes of WRAY16$-$232 span from May 2011 to August 2014.


\begin{table}
\centering
\caption{Photometric and positional properties of WRAY~16-232. Photometric magnitudes are obtained from \citet{Gvaramadze10.1111/j.1365-2966.2010.16496.x}, whereas the distance is taken from the \citet{2023A&A...674A...1G}.}
\renewcommand{\arraystretch}{1.5} 
\setlength{\tabcolsep}{25pt} 
\begin{tabular}{c|c}
\hline
\hline
\textbf{Property} & \textbf{Value} \\
\hline
RA (h:m:s) & 16:43:16 \\
DEC (d:m:s) & $-$46:00:42 \\
distance (kpc) & 2.1 $\pm$ 0.5 \\
\(B\) band & 17.12 \\ 
\(V\) band & 15.66 \\
\(R\) band & 12.92 \\
\(J\) band & 6.26 \\
\(H\) band & 5.08 \\
\(K\) band & 4.21 \\
\hline
\end{tabular}
\label{tab:photo}
\end{table}

\section{Analysis and Results}\label{sec:analysis}

\subsection{Study of the Spitzer IRAC/MIPS images}   \label{sec:irac image analysis}

The Spitzer IRAC 4 and MIPS 1 images show the envelope around WRAY~16-232, which is shown in Fig.~\ref{fig:irs_image}. The MIPS 1 image in the right panel of Fig.~\ref{fig:irs_image} reveals a nearly spherical envelope around the star. The nebulae surrounding LBVs can be bipolar, spherical, or irregular; for example, a spherical shell has been reported around the LBV source S119 \citep{Weis2003A&A...398.1041W}. The spherical shell of WRAY~16-232 is illustrated in \citet{Gvaramadze10.1111/j.1365-2966.2010.16496.x}, although its internal features were not examined in detail in that study. Based on the MIPS 1 image, we estimate the envelope to have a radius of 1.2$\arcmin$, corresponding to 0.70 pc. This falls well within the typical range of LBV dust shell radii, which often span from 0.1 to 2 pc \citep{Weis2020Galax...8...20W}. This shell appears relatively bright and well-defined, suggesting a concentration of dust that is being heated by the central source and re-radiating strongly at 24 $\mu$m. Several brighter arcs or knots of emission trace out an inhomogeneous distribution of material, implying that the dust is clumpy or shaped by localized outflows and wind-wind interactions of faster and slower winds during a S Dor cycle \citep{1999LNP...523...62N}. The extended tail-like structure observed toward the south does not appear to be part of the spherical shell, but rather belongs to a larger, more diffuse region, as confirmed through visual inspection.

The IRAC 4 image has a higher resolution than MIPS 1 and they highlight the clumpy and arc-like denser structures. The 8 $\mu$m image shows warmer dust and potentially traces regions more influenced by PAH emission \citep{Deharveng2010A&A...523A...6D,Arun2021MNRAS.507..267A}, within the main shell. These regions may mark zones where dust composition or temperature differs from the shell’s dominant 24 $\mu$m emission. The arc-like structures appear to converge or point toward the central star, suggesting they may be shaped by directed outflows or strong stellar winds. The “heads” of these structures likely trace regions of denser dust that have been compressed or sculpted by these forces. The IRAC 4 image shows a more concentrated emission in the head structure, indicating the presence or formation of PAHs at an inner radius $<$ 0.7 pc. Overall, the image data depict a rich, layered envelope whose morphology appears to be shaped by multiple mass-loss episodes and complex wind dynamics associated with an evolved massive star. These characteristics offer insights into the star’s LBV nature.

\subsection{MIR Spectral Properties of WRAY~16-232 and Comparison with other LBVs}
\begin{figure*}
    \includegraphics[width=0.75\columnwidth]{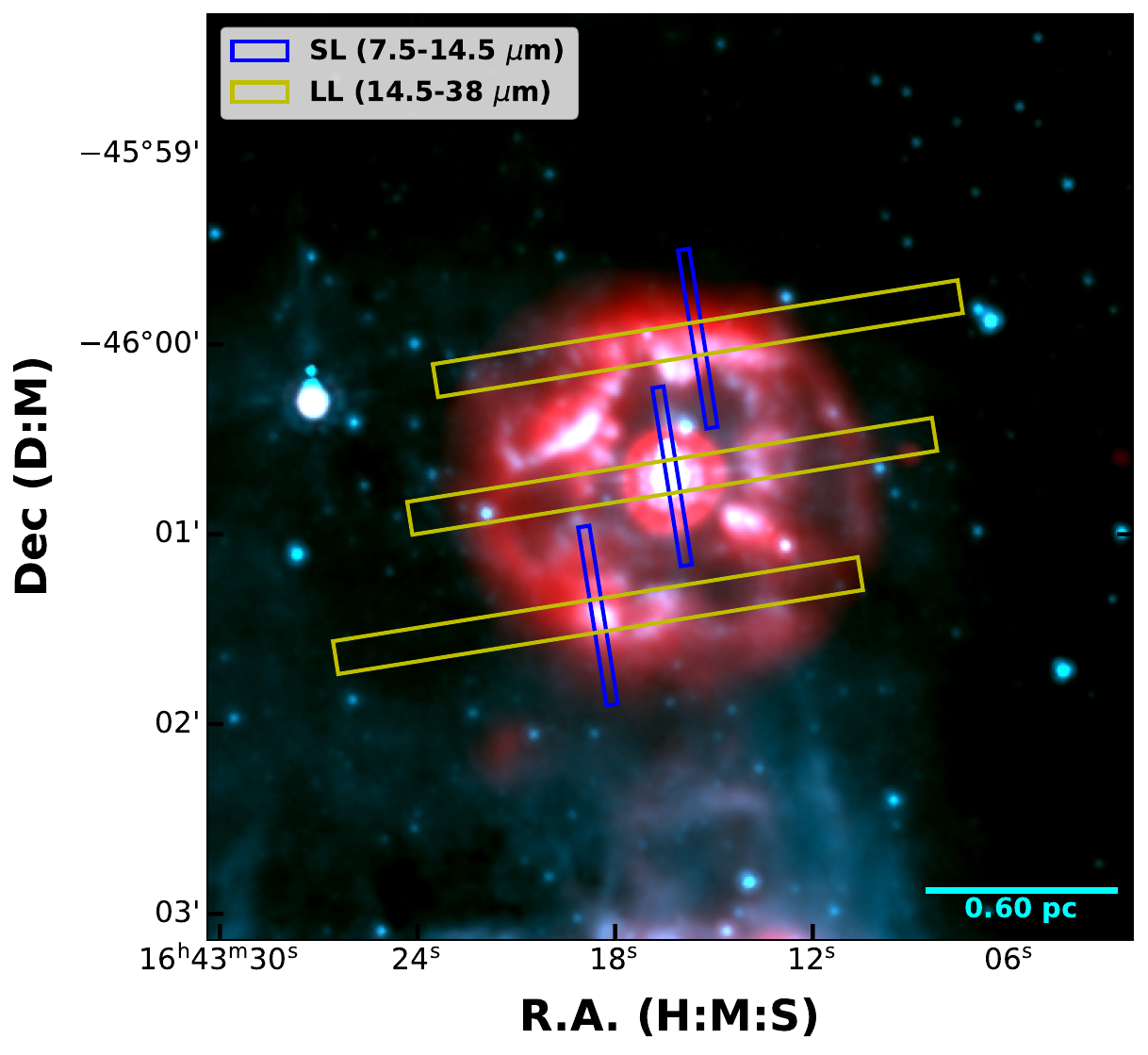}
    \includegraphics[width=1.25\columnwidth]{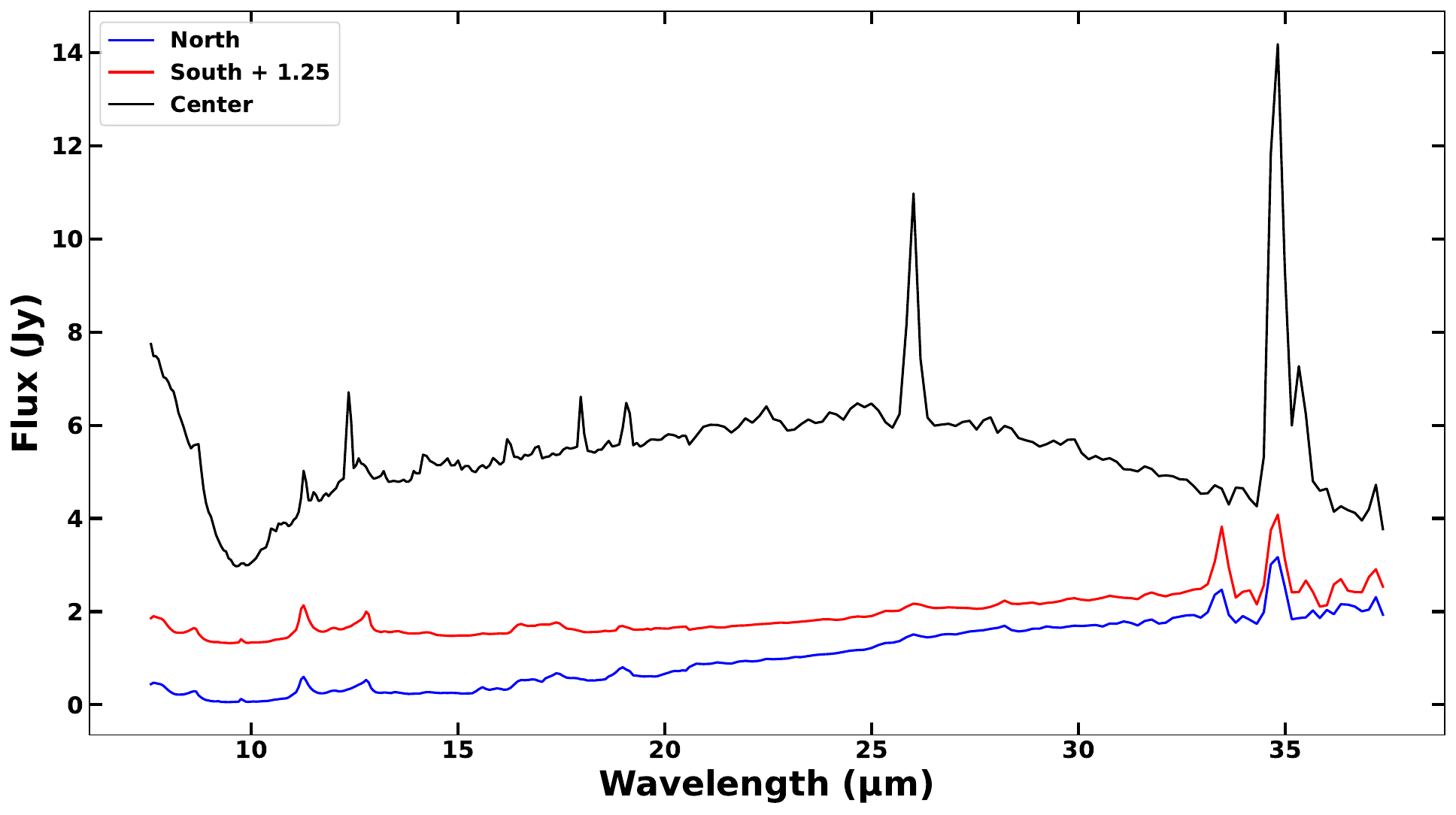}
    \caption{\textbf{Left:} Spitzer IRAC MIPS false color image (8 $\mu$m (IRAC 4) - turquoise and 24 $\mu$m (MIPS 1) - red) is shown with slit orientation of three pointing from Spitzer IRS Short Low (SL) and Short High (SH). \textbf{Right:} Spitzer IRS spectrum of the three regions. The south pointing flux is increased for visibility of the figure.}
    \label{fig:irs_spectra}
\end{figure*}

\begin{figure*}
    \includegraphics[width=1.25\columnwidth]{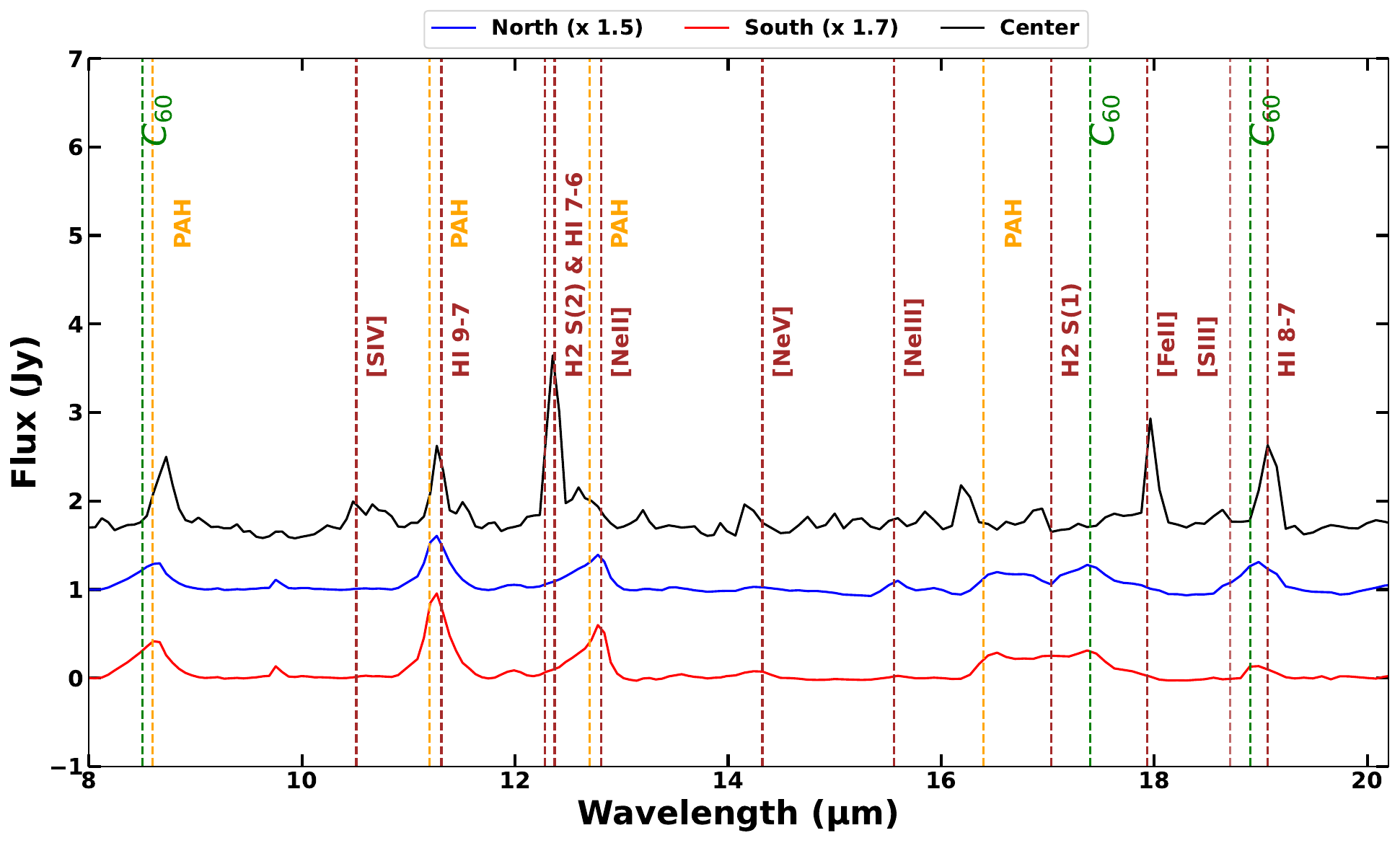}
    \includegraphics[width=0.74\columnwidth]{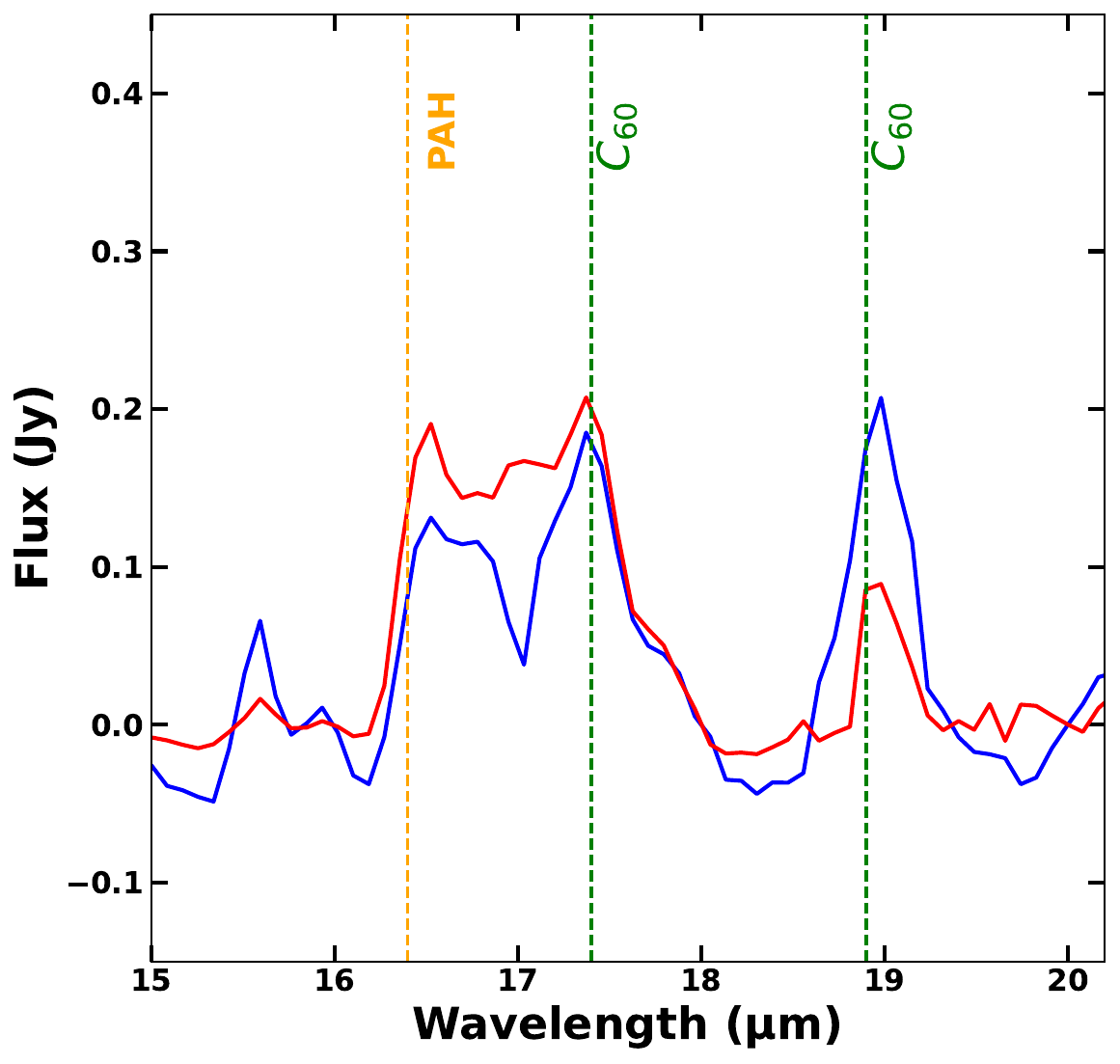}
    \caption{Continuum-subtracted Spitzer/IRS spectra of the center (black), north (blue), and south (red) regions of WRAY~16-232 (left), offset vertically for clarity. Vertical dashed lines mark prominent emission lines and PAH/fullerene features, with labels denoting their identifications. Each spectrum is scaled as indicated in the legend to highlight variations in line strengths among the regions. The zoomed-in region shows the 17.4 and 18.9 $\mu$m C$_{60}$ features identified in the northern and southern parts of the envelope (right). }
    \label{fig:irs_spectra_contin}
\end{figure*}

Continuum subtraction for the LR (R~$\sim$ 60) spectra of WRAY~16-232--covering the central source and two envelope positions to the north and south (Fig.~\ref{fig:irs_spectra}) was performed using a spline-fitting method similar to that employed by \citet{Seok_2017} and \citet{10.1093/mnras/stad1511}. Anchor points were selected in the 7–36 $\mu$m range, avoiding strong emission lines and PAH features and were modeled separately for the central and envelope spectra. For the central source, anchor points were placed at 7.6, 8.0, 8.3, 9.25, 10.3, 11.7, 13.0, 13.7, 13.9, 14.1, 14.4, 14.6, 15.35, 16.0, 18.3, 19.8, 20.3, 21.5, 23.4, 27.0, 28.5, 30.0, 32.0, 33.0, 34.1, 36.1, and 36.8 $\mu$m, while for the northern and southern envelopes, they were set at 7.6, 8.0, 8.1, 9.25, 10.3, 11.7, 13.1, 14.5, 15.35, 16.2, 18.3, 19.8, 20.0, 22.0, 24.0, 28.0, 30.0, 32.0, 33.0, 33.8, 35.2, and 36.5 $\mu$m. A cubic spline fit to these points modeled the continuum, which was then subtracted to isolate emission features.
The continuum-subtracted spectra, which is shown in Fig.~\ref{fig:irs_spectra_contin}, exhibit distinct variations across the three pointings, particularly in continuum morphology and emission features. The key similarities and differences are summarized below and further analyzed in the context of MIR spectral characteristics of other known LBVs.

\paragraph*{(i) Dust Continuum and Silicate Absorption:} The spectrum corresponding to the central pointing is characterized by a rising continuum beyond the silicate absorption feature at 9.7 $\mu$m. This is similar to what is observed in the Galactic cLBV G79.29+0.46 \citep{Jiménez-Esteban_2010}. The absorption points to optically thick and cooler dust temperatures in the line of sight. An $A_{\rm V}$ value of 12 mag measured using the optical depth of 9.7 $\mu$m silicate absorption for WRAY~16-232 supports an optically thick line of sight \citep{Nowak_2014}. In contrast, this silicate feature appears in emission in many other LBVs, such as HR Car \citep{2009ApJ...694..697U}, HD 168625 \citep{Umana2010ApJ...718.1036U} and WRA 751 \citep{2000A&A...356..501V}.

Interestingly, the envelope spectra of WRAY 16‑232 show flatter mid‑IR continua and weak or absent silicate absorption, suggesting either reduced silicate dust or different excitation conditions farther from the central star. Notably, the northern envelope slope beyond 15 $\mu$m is steeper than the southern one, which may reflect variations in dust column density or temperature along different viewing directions. 

\paragraph*{(ii) Ionized and Atomic Fine-Structure Lines:} In the central pointing, we detect a range of ionic lines, such as those of [Si~{\sc ii}] 34.83 $\mu$m and [Fe \textsc{ii}] transitions at 17.96, 25.99, and 35.35 \micron. These lines are characteristic of the presence of some hot, ionized environments, associated with massive stars having strong UV fields. Similar fine-structure lines are also observed in the MIR spectra of known LBVs \citep{Umana2010ApJ...718.1036U} e.g., HR Car \citep{2009ApJ...694..697U}, alongside strong free-free or dust emission. The envelope spectra also feature some faint lines such as [Ne~\textsc{iii}] 15.55 \micron, [Fe~\textsc{ii}] 25.99 \micron, [S~\textsc{iii}] 33.49 $\mu$m. This reduction in line strength is likely due to decreased excitation or density with increasing distance from the central star.

\paragraph*{(iii) Hydrogen Recombination and Helium Lines:} Hydrogen recombination lines (e.g., H~\textsc{i} at 8.76, 12.37, 16.88, 19.06, 27.8 \micron) and He~\textsc{ii} lines (10.47, 13.13, 22.18 \micron) are detected across all three pointings, with the strongest and most numerous features observed in the central source. Strong hydrogenic emission lines are a key mid-IR signature of massive stars with substantial ionizing radiation \citep{2017MNRAS.470.2710M}.  Similar features have been identified in LBVs such as P Cygni and AG Car, often accompanied by prominent P Cygni profiles in the optical or near-IR \citep{1997A&A...326.1117N}. Their detection in both the stars and its surrounding envelope suggests the presence of an extended ionized region shaped by strong stellar winds.

\paragraph*{(iv) Molecular Hydrogen and PAHs:} Both envelope pointings exhibit H\textsubscript{2} rotational lines, including S(0) at 28.22 \micron\ and S(3) at 9.66 \micron. Additionally, the S(2) line at 12.28 \micron\ is detected in the spectrum of the central object. Pure rotational lines of H\textsubscript{2} have been identified in only a few LBVs, such as the candidate LBVs HDE 316285 \citep{2008IAUS..250..361M} and G79.29+0.46 \citep{Jiménez-Esteban_2010}. Alongside the H\textsubscript{2} lines, we also report the detection of prominent PAH bands at 8.6, 11.2, 12.7, and possibly 16.4 \micron, consistent with previous detections in other LBVs and cLBVs \citep{1997ASPC..120..322S, 1999A&A...341L..67V, 2000A&A...356..501V}.

The strong PAH emission in the envelopes of WRAY~16-232 suggests the presence of carbon-rich ejecta, likely processed in the extended dust shell, possibly a remnant of earlier mass-loss episodes. Additionally, the presence of 16.4 $\mu$m and 17.4 $\mu$m PAH features, later found to be contaminated by C$_{60}$ and attributed to larger PAHs containing 50-130 carbon atoms, has not been previously reported in evolved stars and LBVs \citep{Boersma2010A&A...511A..32B}. Their presence indicate that the envelope around WRAY~16-232 has optimal environment for forming larger PAH molecules. In contrast, the absence or suppression of PAH features along the line of sight toward the central star may be attributed to a hotter dust environment or silicate-dominated material, which does not efficiently form the PAHs.

\paragraph*{(v) Fullerene Signatures:} As shown in the right panel of Fig.~\ref{fig:irs_spectra_contin}, the envelope spectra of WRAY~16-232 reveal distinct C\textsubscript{60} features at 17.4 \micron\ and 18.9 \micron, whereas we did not notice these lines in the central source. This marks the first reported detection of fullerenes in a cLBV envelope, given that fullerene emission in high-mass star ejecta has not been reported to date and has, thus far, been more commonly observed in PN (e.g., \citealp{doi:10.1126/science.1192035,García-Hernández_2012}). Since LBVs undergo episodic mass loss and dust formation in harsh radiative environments, the presence of C\textsubscript{60} in WRAY~16-232 provides a unique window into carbon processing under intense UV fields. The 16.4 \micron\ PAH emission is notably stronger in the southern spectrum, and the 17.4 \micron\ feature also shows enhanced intensity there, indicating that the feature has a PAH contribution along with the fullerene as mentioned in the literature. 

The mid-IR spectral properties of WRAY~16-232 exhibit key LBV characteristics, including a strong dust continuum, presence of fine-structure lines [Fe~\textsc{ii}], [Ne~\textsc{iii}] and [S~\textsc{iii}] typical of ionized nebulae, and a chemically diverse circumstellar environment. The combination of silicate absorption in the central source with PAH and fullerene emission in the envelope points to a complex, multi-stage mass-loss history, potentially influenced by  episodic ejections at different evolutionary phases or binarity. The detection of C\textsubscript{60} further highlights the role of UV-driven processes in shaping the circumstellar medium. These findings suggest that WRAY~16-232 is an evolved massive star undergoing active dust processing, with its intricate chemical composition and mass-loss patterns warranting further investigation.

\subsection{Comparison with other C\textsubscript{60} Sources}

\begin{figure}
    \includegraphics[width=1\columnwidth]{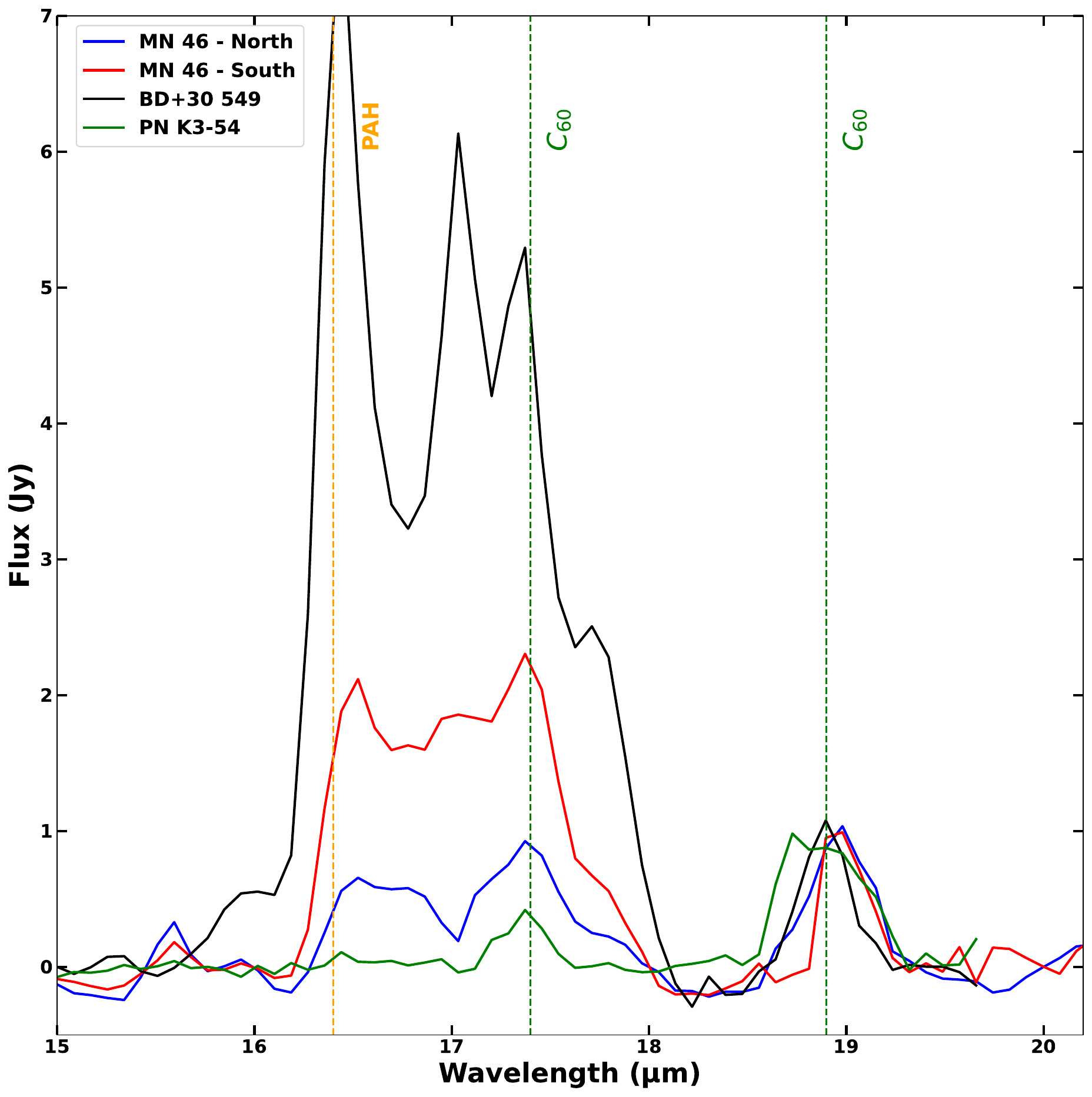}
    \caption{Continuum-subtracted Spitzer/IRS spectra of WRAY~16-232 (north and south), BD+30 549 (RNe), and PN K3-54, normalized to unity at the 18.9,\micron\ C\textsubscript{60} feature. The 17.4 and 18.9 \micron\ fullerene bands are detected in all sources, but their relative strengths vary depending on the physical conditions of the local environment.}
    \label{fig:irs_spectra_comparison}
\end{figure}

We compare the continuum-subtracted spectra of WRAY~16-232 (north and south regions) with two well-known fullerene-rich sources: BD+30 549, a RNe and PN K3-54 \citep{10.1093/mnras/stt2070, 10.1093/mnras/stad1511} in Fig.~\ref{fig:irs_spectra_comparison}. All spectra are normalized to unity at the 18.9 \micron\ C\textsubscript{60} feature to facilitate direct comparison.

The 16.4 \micron\ PAH emission varies significantly among the sources. BD+30 549, which is associated with a RNe known for strong PAH emission, exhibits an enhanced 16.4 \micron\ PAH feature along with a higher level of PAH contamination in the 17.4 \micron\ region. In contrast, the PN K3-54 spectrum is nearly devoid of PAH emission, showing only the pure C\textsubscript{60} features at 17.4 and 18.9 \micron. Notably, in the PN spectrum, the 17.4 \micron\ feature is significantly weaker than the 18.9 \micron\ feature, which is characteristic of a fullerene-dominated environment \citep{García-Hernández_2010}.

WRAY~16-232 presents an intermediate spectrum between these two extremes. While PAH emission is evident, particularly in the southern envelope, the 16.4 \micron\ PAH feature is not as intense as in BD+30 549, suggesting a less PAH-rich environment. However, the relative intensity of the 17.4 \micron\ feature is elevated in the south, indicating that PAH contamination is contributing to this band, unlike in PN K3-54, where the 17.4 \micron\ feature is purely due to C\textsubscript{60}. This suggests that fullerene formation in WRAY~16-232 occurs in an environment with active PAH processing, distinguishing it from the more chemically evolved PN K3-54, where PAHs are largely absent. The comparison highlights how LBV ejecta can provide conditions for fullerene formation that are distinct from both RNe and PN.

\begin{figure*}
    \includegraphics[width=1.35\columnwidth]{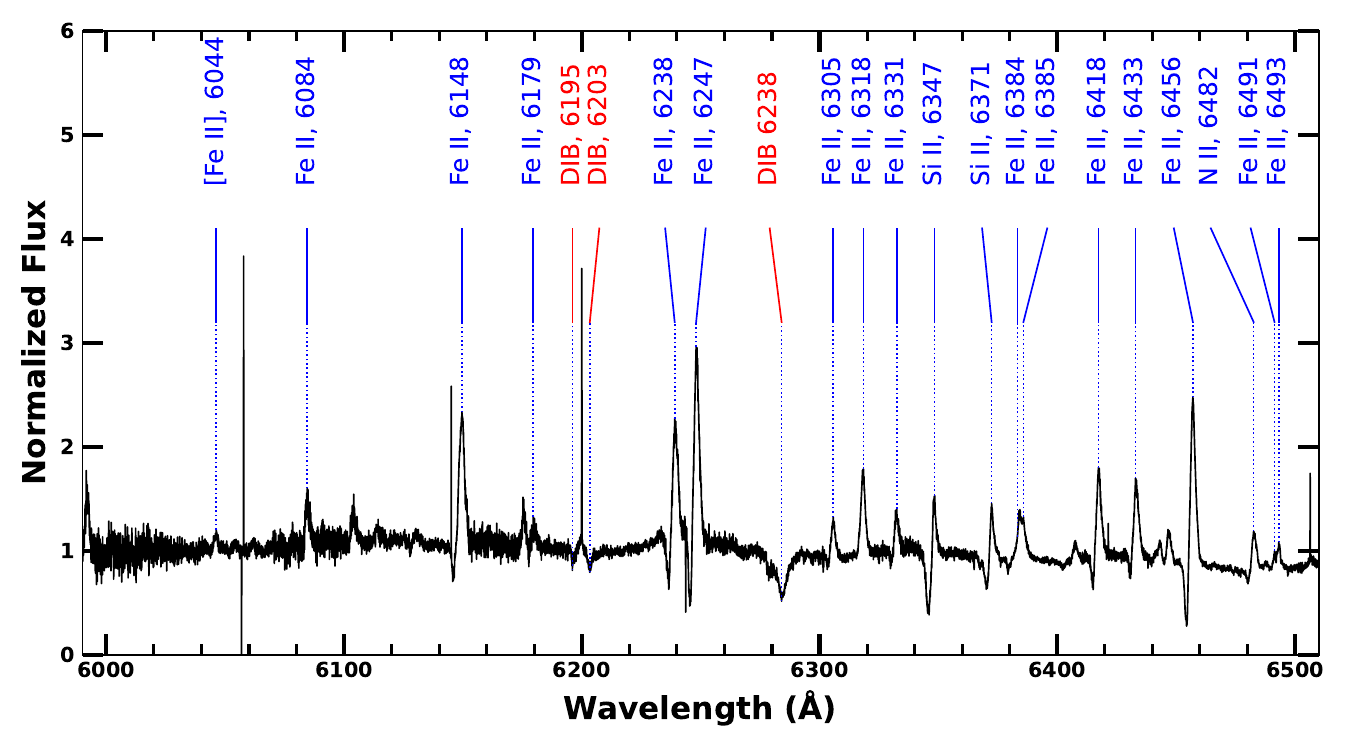}
    \includegraphics[width=0.7\columnwidth]{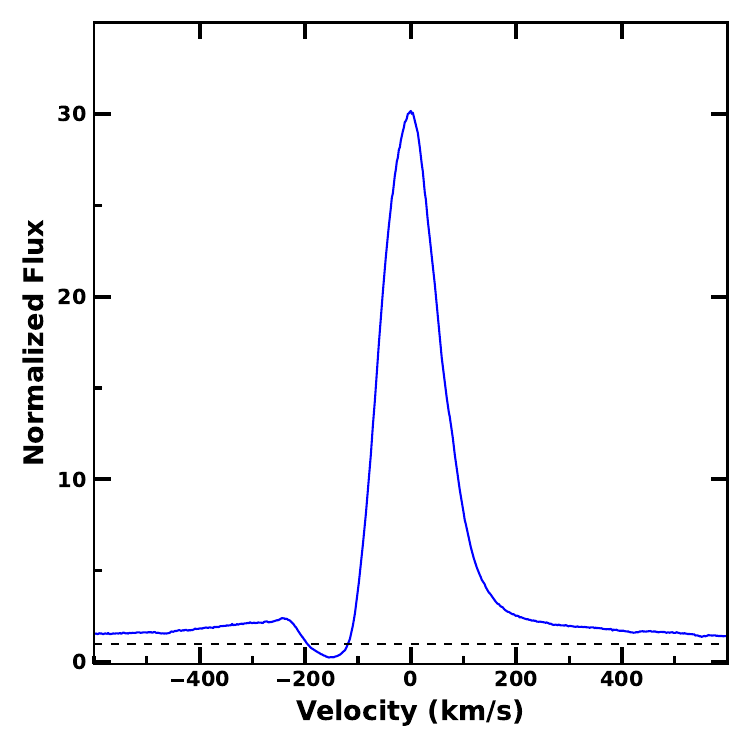}
    \caption{Portion of SALT HRS spectra of WRAY~16-232 in the left panel (obtained on 16 May 2016). The thin lines in the region are likely caused by cosmic ray hits or instrumental artifacts. The right panel depicts the H$\alpha$ line in WRAY~16-232, where the blue edge of the absorption component roughly indicates the terminal velocity.}
    \label{fig:spectra}
\end{figure*}

\subsection{Analysis of SALT HRS Spectra} \label{sec: optical}


By analyzing the SALT HRS spectra of WRAY~16-232, we identified a total of 109 emission lines, with P~Cygni morphology being the most prevalent. This indicates the presence of an expanding circumstellar shell and substantial mass loss from the star \citep{Rottenberg10.1093/mnras/112.2.125}. Additionally, the near absence of stellar photospheric absorption lines suggests that the spectral continuum is primarily dominated by stellar winds.

The H$\alpha$ line exhibits a P Cygni profile (Fig.~\ref{fig:spectra}) with very broad wings, indicative of regions with significant electron scattering \citep{Bernat_1978}. In case of classical Be stars, the H$\alpha$ line normally does not show
the P Cygni profile regardless of the environment \citep{Rivinius2013A&ARv..21...69R, Banerjee_2022} which completely rules out the SIMBAD classification of WRAY~16-232 as a Be star. Additionally, all Paschen series lines display P Cygni profiles. Weak P Cygni profiles are observed in He~{\sc i} $\lambda\lambda$5876, 6679, 7065, and 8582, characterized by very broad absorption components that are significantly shifted from their rest wavelengths, suggesting the presence of a high-velocity region. The spectrum also features a forest of Fe {\sc~ii} lines, some appearing in pure emission and others with P Cygni profiles. Prominent Fe{\sc~ii} P Cygni lines include $\lambda\lambda$6149, 6456, and 7711. Additionally, three [Fe~{\sc ii}] forbidden lines were detected.

The analysis of the HRS spectra reveals the presence of singly ionized lines of N~{\sc ii}, Si~{\sc ii}, Al~{\sc ii}, Ca~{\sc ii}, and Mg~{\sc ii}. Among these, the detected Ca~{\sc ii} lines include the well-known Ca~{\sc ii} triplet ($\lambda\lambda$8498, 8542, 8662), which is blended with Paschen lines (P16, P15, and P13, respectively). The presence of the Ca~{\sc ii} triplet suggests a cooler region (T $\sim$ 5000 K; \citealp{2000A&A...353..666C}) within the circumstellar environment. Additionally, several N~{\sc i} lines were identified, which are characteristic features commonly observed in many LBVs (\citealp{1982ApJ...254L..47D}). O~{\sc i} lines, including the triplet at $\lambda$7775 and $\lambda$8446 (blended with an Fe~{\sc ii} line), were also detected.

Moreover, several fine-structure lines are identified in the HRS spectra of WRAY~16-232 as well. These include multiple [Fe~{\sc ii}] lines, which are typical features of LBVs, as well as the [N~{\sc ii}] $\lambda\lambda$5755 and 6584 nebular lines. The [N~{\sc ii}] $\lambda$6584 line is superimposed on the broad emission wings of H$\alpha$. The [N~{\sc ii}] line also appears to be blended with the C~{\sc ii} 6582 feature, which is commonly observed in other LBVs, like P~Cygni. The [Fe~{\sc ii}] $\lambda$7155 and [N~{\sc ii}] $\lambda$5755 lines, if exhibiting flat-topped profiles, could potentially be used to calculate the terminal velocity ($v_{\infty}$; \citealp{1991A&A...244..467S}). No such line structures were observed in the spectra, precluding direct velocity measurement. Nevertheless, by tracing the blue edge of the absorption components, an upper limit for the terminal velocity was estimated to be $v_{\infty} \approx 244$ km s$^{-1}$. This method has also been widely adopted by other authors, for example, \citet{Kniazev_2016}. Our estimated $v_{\infty}$ closely matches that of the prototype Galactic LBV AG Car during its quiescent phase, which was reported by \citet{Groh_2009} to be $270 \pm 50$ km s$^{-1}$. This similarity suggests that WRAY~16-232 was likely in a quiescent phase at the time spectra were obtained, i.e. during May 2016. It is also worth noting that a lower terminal velocity of 225 km s$^{-1}$ was estimated by \citet{2001A&A...375...54S} during AG Car's visual minimum phase, although they acknowledged that their estimate, based on a $\beta$-law velocity profile, likely underestimates the true terminal velocity. To better contextualise the $v_{\infty}$ value of WRAY~16-232, the terminal wind velocities of other LBVs in their quiescent phases are as follows: P~Cygni - 310~km s$^{-1}$~\citep{Lamers1985}, HD~168625 - 350~$\pm$~100~km s$^{-1}$~\citep{Mahy2016}, Romano's Star - 265~km s$^{-1}$~\citep{Clark2012}, and [GKF~2010]~MN48 - 152~$\pm$~5~km s$^{-1}$~\citep{Kniazev_2016}. A $v_{\infty}$ of approximately 244 km s$^{-1}$ is consistent with the observation that LBV winds are slower compared to those of other supergiants \citep{Vink_2018}. The right panel in Fig.~\ref{fig:spectra} illustrates the H$\alpha$ line in the WRAY~16-232 spectrum.

Numerous Diffuse Interstellar Bands (DIBs) were observed in the spectrum, including $\lambda\lambda$5780, 5797, 5849, blended 6195 and 6196, 6203, 6613, 6660, 6699, 6729, 8621, and a broad, blended feature spanning $\lambda\lambda$6276–6288. The identification of DIBs in our spectra was made by comparing with the features reported in \citet{1994A&AS..106...39J,2006A&A...448..221S,2022RAA....22h5007W}. Additionally, a broad interstellar absorption band corresponding to Na D1 and D2 is present.

Measuring the extinction parameter ($A_{\rm V}$) using the strength of DIBs is a common approach employed in numerous studies \citep{Carvalho_2022, Nidhi10.1093/mnras/stad2067, Ma2024A&A...691A.282M}. We calculated the $A_{\rm V}$ using the DIB strength of the 6613 \(\text{\AA}\) line. The equivalent width (EW) of the DIB was related to the reddening value \(E(B-V)\) using the equation:
\vspace{-0.1em}
\begin{equation}
E(B-V) = a + b \cdot \text{EW}_{\text{DIB}}, \label{eq:ebv_dib}
\end{equation}

\noindent where the coefficients \(a = 1.96 \times 10^{-2}\) and \(b = 4.63 \times 10^{-3}\). The Equation~\ref{eq:ebv_dib} is adopted from \citet{Friedman_2011}. The extinction, $A_{\rm V}$ =  $R_{\rm V}$ $\cdot$ E(B-V), assuming an total-to-selective extinction of \(R_{\rm V} = 3.1\) (for the Milky Way) we determined $A_{\rm V}$ = 0.0664. The DIB based $A_{\rm V}$ values are only reliable for measuring interstellar (line-of-sight) extinction, but they do not account for extinction caused by the envelope around a LBV \citep{Klochkova2022ARep...66..998K}. One other caveat is that the galactic $R_{\rm V}$ value would change for such environments due to the envelope in the line of sight, as indicated by the silicate absorption. For example, in the case of LBV HR Car, the measured $R_{\rm V}$ is 4.3 $\pm$ 0.1, indicating much larger grain sizes in its circumstellar environment compared to the average interstellar value \citep{Mehner2021A&A...655A..33M}. Hence, in order to resolve the ambiguity, we have adopted the $A_{\rm V}$ from \citet{Nowak_2014} for the present study.

\citet{2000AJ....119.2214M} and references therein argued that the presence of a dense forest of [Fe~{\sc ii}] lines indicates that the LBV is in its visual minimum state, during which the temperature of LBVs ranges from approximately 12,000 K to 30,000 K \citep{2007ASSL..342.....K}. In the optical spectra of WRAY~16-232, we observe a few [Fe~\textsc{ii}] lines, though not a dense forest. In contrast, there is a prominent forest of Fe~\textsc{ii} lines. Coupled with the absence of higher ionization features such as Fe~\textsc{iii}, this suggests the presence of relatively low-temperature regions in the vicinity of WRAY~16-232. Fe~\textsc{iii} lines are generally expected during visual minimum or quiescent phases, when LBV temperatures are high (12,000–30,000 K). Their absence may be due to the high optical depth reported by \cite{Nowak_2014}, which could have obscured these lines.

LBVs can exhibit episodic mass loss, leading to variations in wind velocity at different distances from the central star. A common spectroscopic method for determining LBV wind velocities involves using the Doppler shift formula.

We computed wind velocities for all unblended P Cygni lines in the optical spectrum of WRAY~16-232 (Appendix~\ref{line_velocity}). The velocity varies from 75 – 160 km s$^{-1}$ in various species (Fig.~\ref{fig:vel_box_plot}). The H$\alpha$ and He~\textsc{i} lines exhibit the highest velocities (140 – 160 km s$^{-1}$) due to their strongly blue-shifted absorption components, indicating that these transitions might be occurring in the faster moving inner regions of the envelope. In contrast, the N~\textsc{i}, Si~\textsc{ii}, and Fe~\textsc{ii} lines show lower wind velocities of 80 – 120 km s$^{-1}$, potentially forming in denser, slower-moving layers closer to the base of the wind or within the inner regions of the layered envelope. We also computed the average velocities for H~\textsc{i}, He~\textsc{i}, N~\textsc{i}, and Fe~\textsc{ii}, which are approximately 97, 161, 86, and 105 km s$^{-1}$, respectively. The variation in average velocities among these different spectral species highlights the multi-layered ionisation structure of the WRAY~16-232 envelope. The wind-wind interactions of faster and slower winds can lead to the formation of bright arcs and knots found in Spitzer images \citep{Weis2020Galax...8...20W}. 

\subsection{Light Curve Analysis of WRAY~16-232}

Photometric variability, known as the S Dor cycle, is a key observational feature used to classify a star as an LBV \citep{2001A&A...366..508V}. Photometric studies of AG Car and S Doradus, both bona fide LBVs, have identified two distinct S Dor cycles: a shorter cycle lasting less than 10 years and a longer cycle exceeding 20 years \citep{VanGenderen1997A&A...318...81V}. Notably, no SD cycles have been observed with durations between 10 and 20 years.

\begin{figure*}
    \includegraphics[width=\columnwidth]{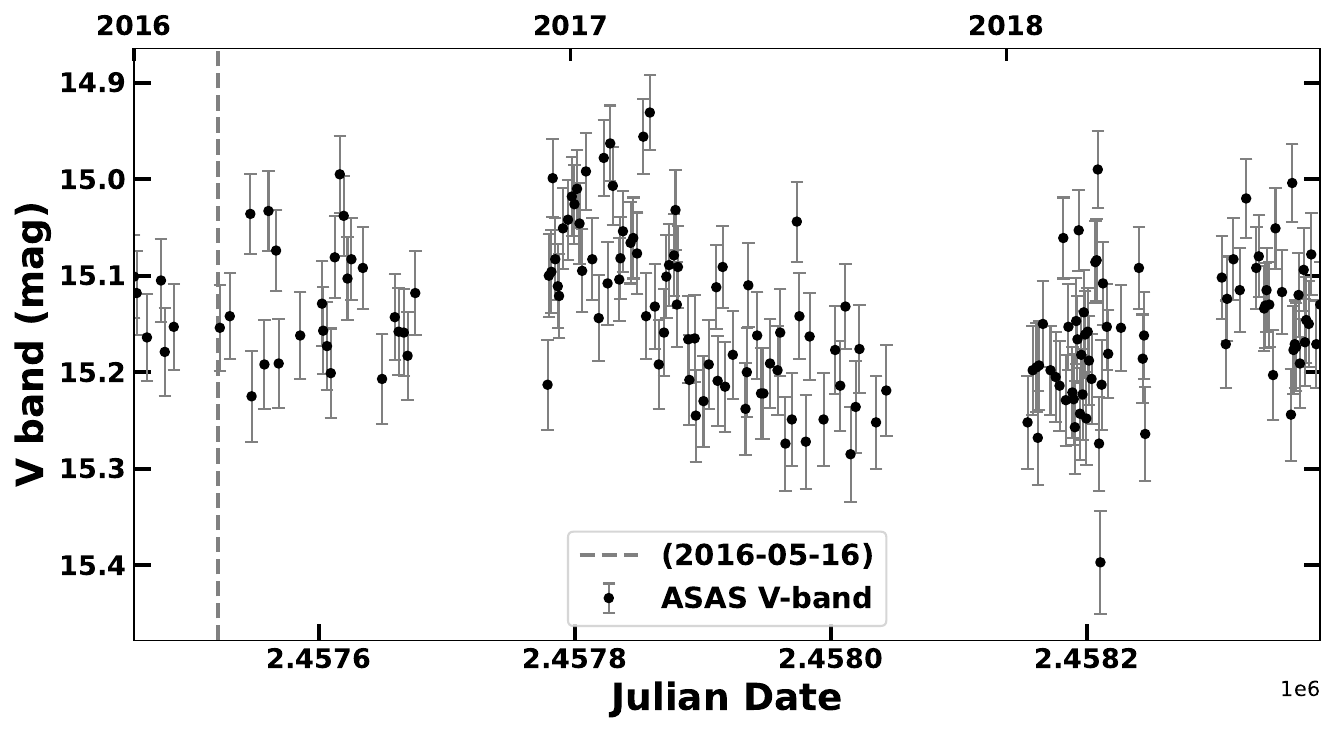}
    \includegraphics[width=\columnwidth]{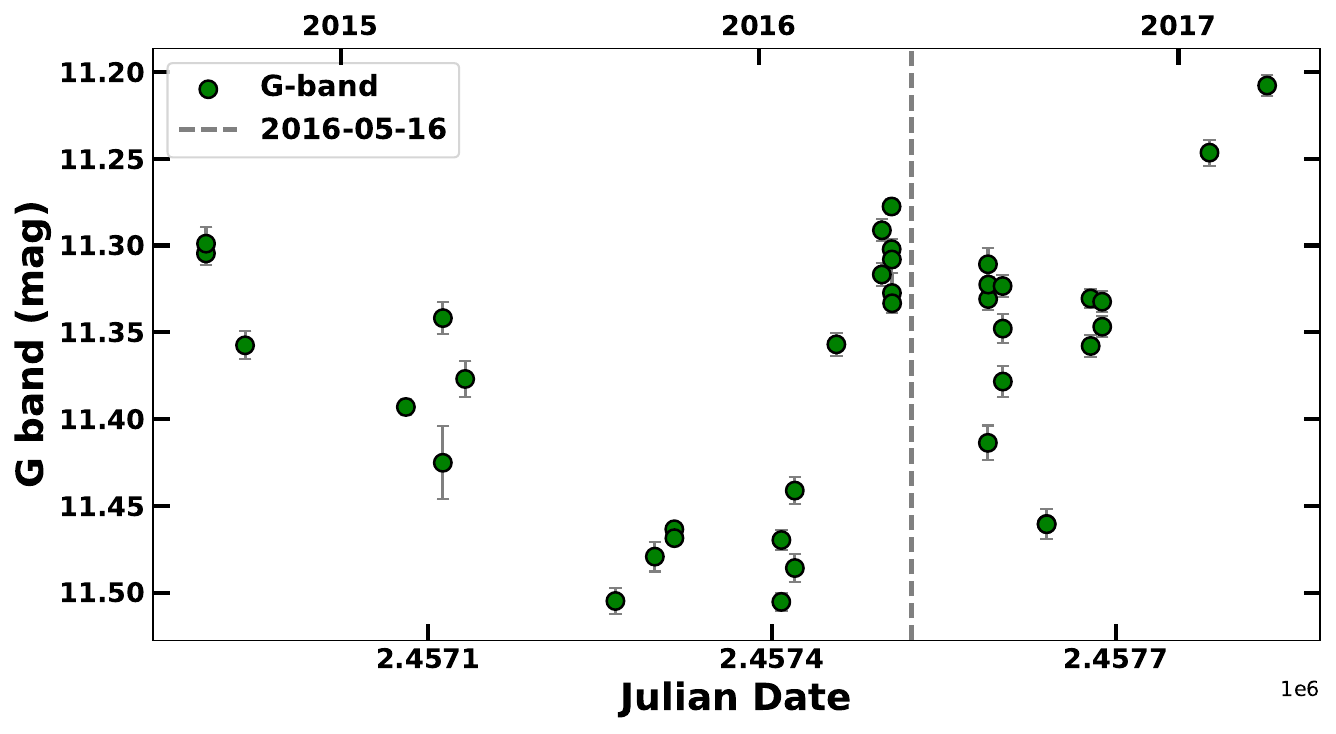}
    \includegraphics[width=\columnwidth]{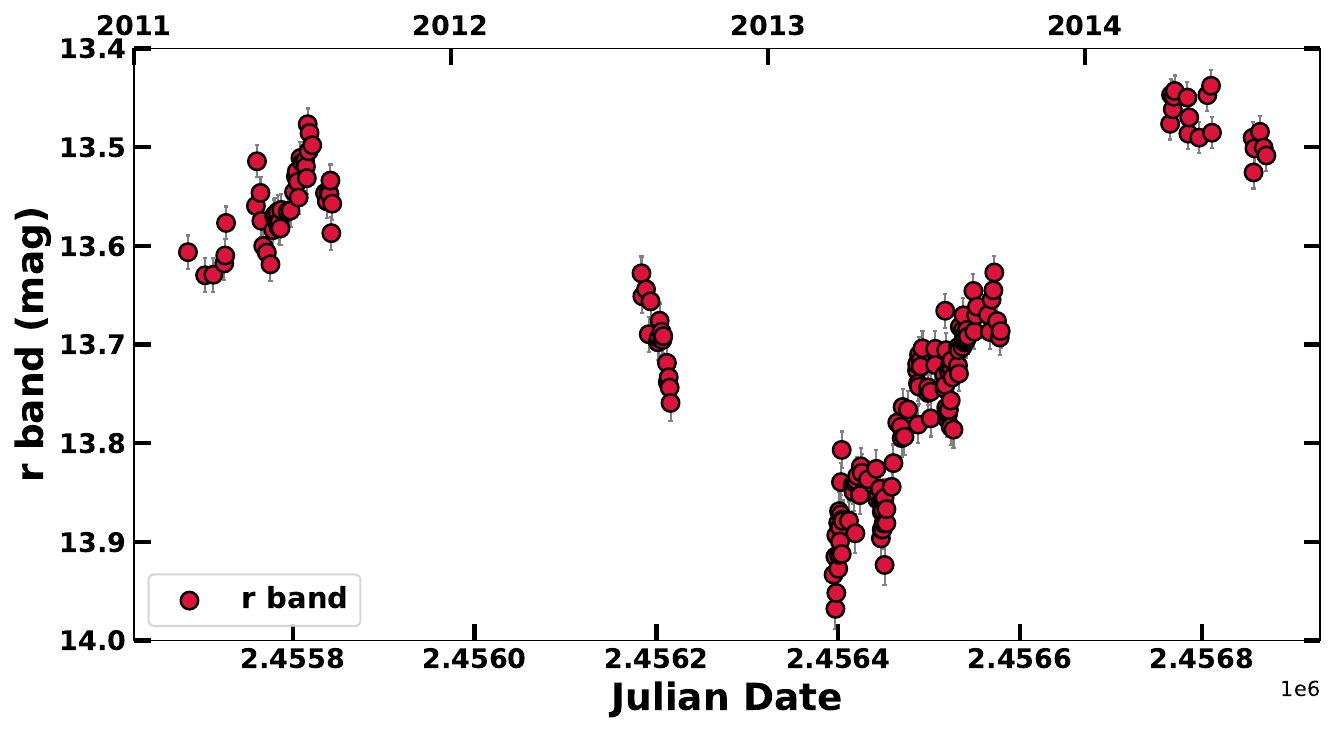}
    \includegraphics[width=\columnwidth]{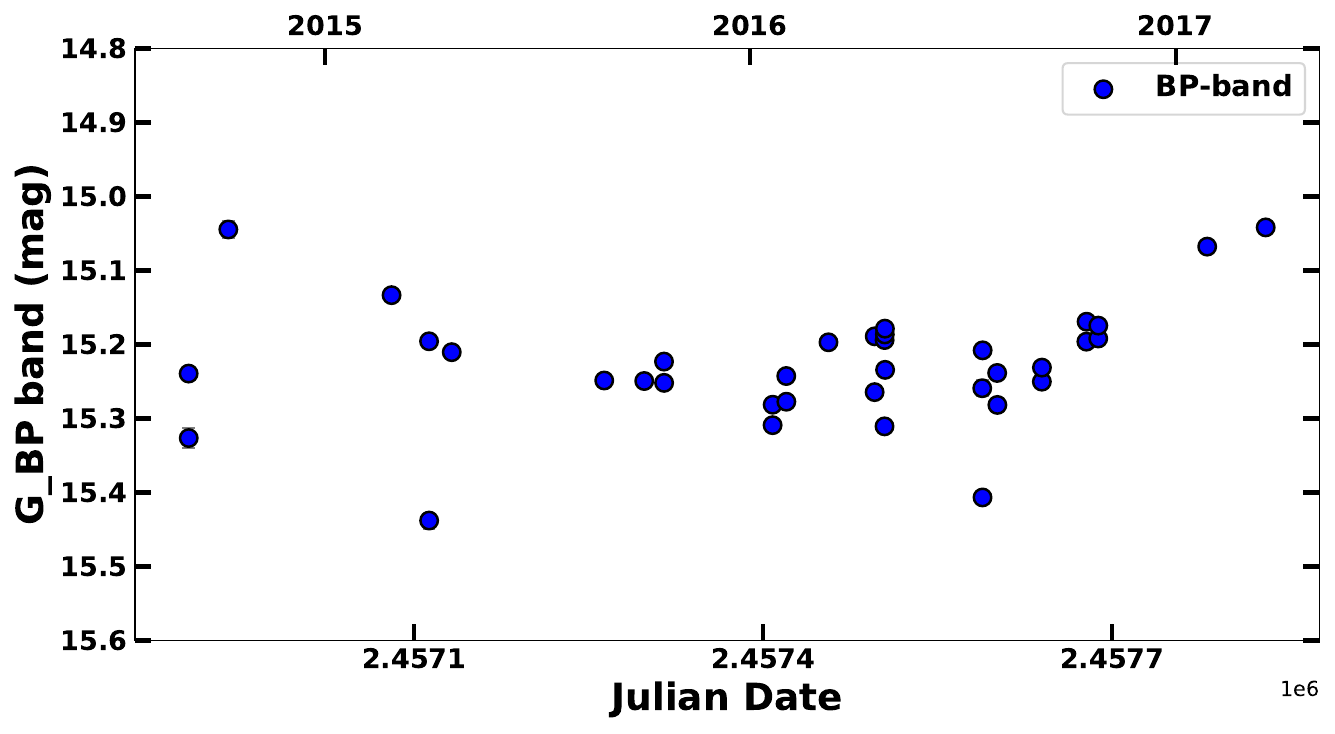}
    \includegraphics[width=\columnwidth]{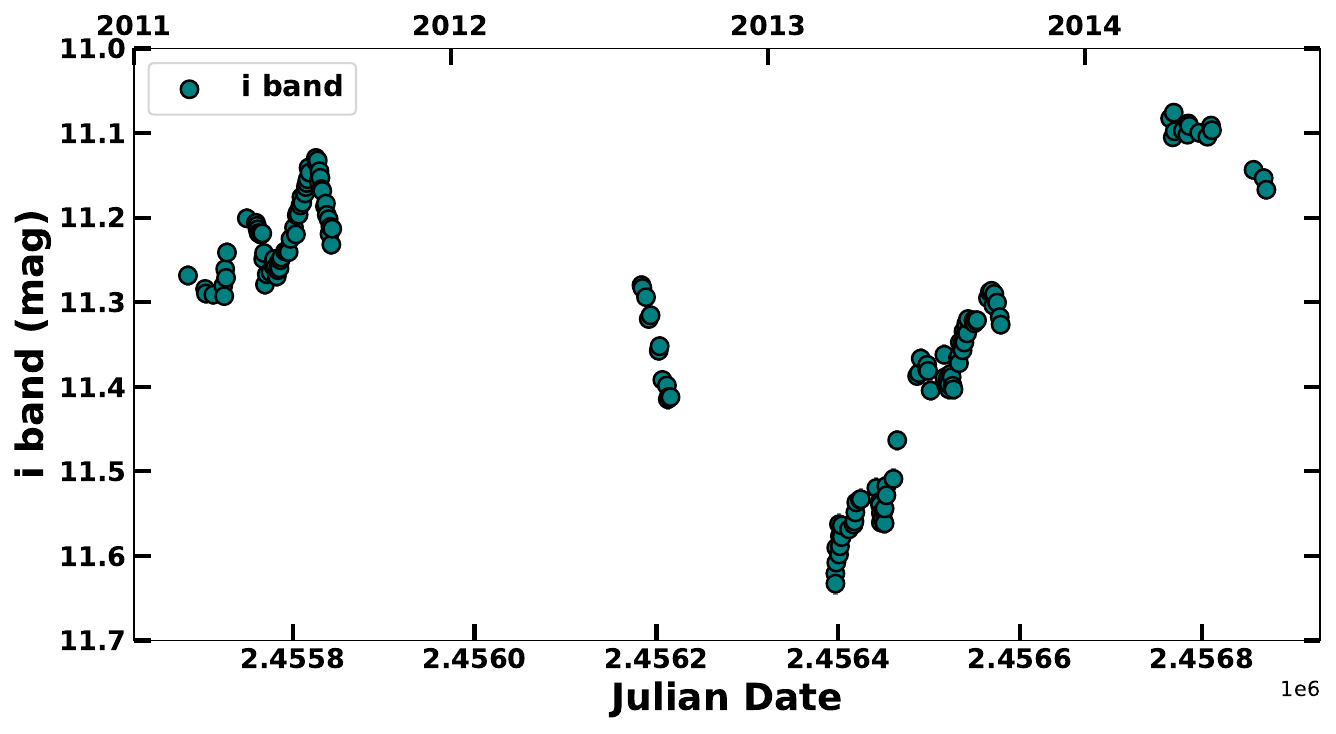}
    \includegraphics[width=\columnwidth]{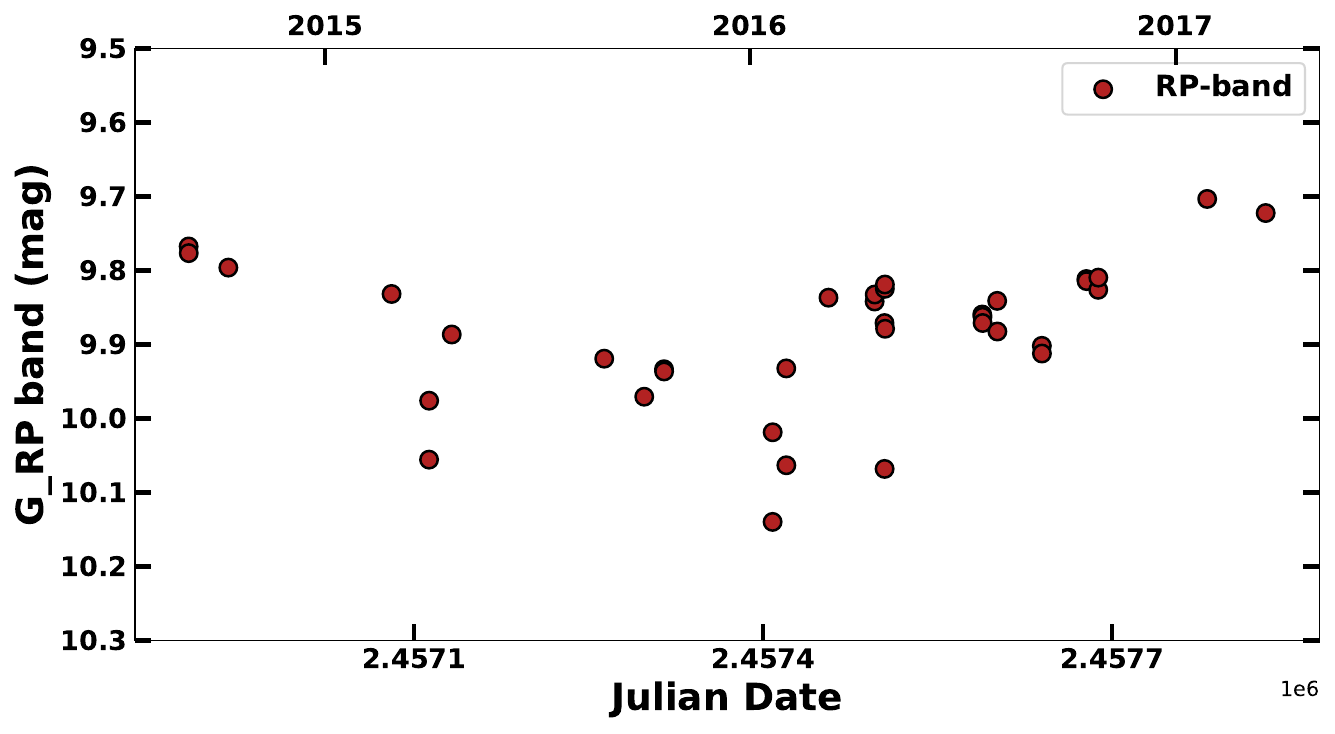}
    \includegraphics[width=1.9\columnwidth]{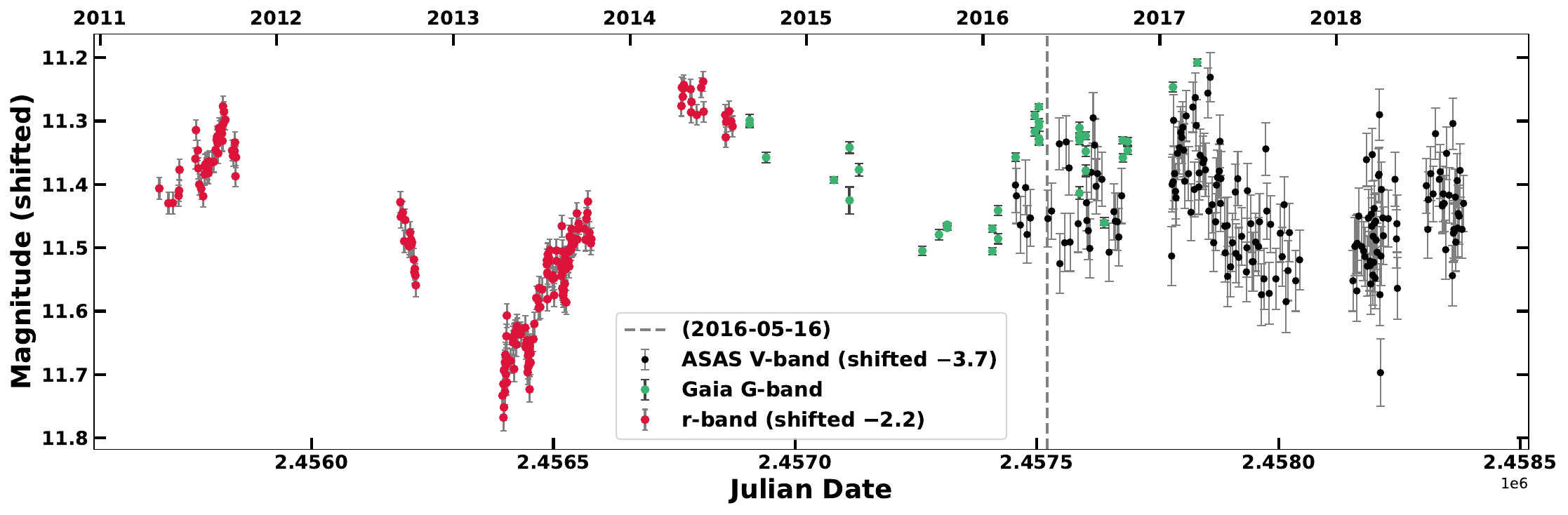}
    \caption{Multi-band photometric variability of WRAY~16-232 from different surveys and filters. \textbf{Top row:} Left panel shows the \textit{ASAS-SN} V-band light curve from 2016--2018; right panel shows the \textit{Gaia} G-band light curve from 2015--2017. \textbf{Middle row:} Left panel displays \textit{r}-band magnitudes; right panel shows \textit{Gaia} \textit{BP} magnitudes from 2015--2017. \textbf{Third row:} Left panel presents \textit{i}-band magnitudes; right panel shows \textit{Gaia} \textit{RP} magnitudes. \textbf{Bottom panel:} Combined light curve showing \textit{r}-band (\textit{r}-band offset by $-2.2$~mag), V-band (\textit{ASAS-SN}, offset by $-3.7$~mag) and \textit{Gaia} G-band photometry, with magnitudes shifted for visual comparison. The vertical dashed line marks the epoch when the spectrum was obtained.}
    \label{fig:Light_curves}
\end{figure*}

Fig.~\ref{fig:Light_curves} presents all the light curves we plotted for WRAY~16-232. The \textit{Gaia} G-band light curve exhibits variability of approximately 0.2–0.3 magnitudes over its observational period, while the \textit{ASAS-SN} V-band light curve displays a comparable level of variability during its observations. A gray vertical line indicates the date on which the spectra of WRAY~16-232 were obtained. This level of variability is consistent with the short-term fluctuations commonly observed in LBVs, likely driven by pulsational activity \citep{1998A&A...335..605L}, commonly observed in massive evolved stars.

We have also plotted the \textit{Gaia} BP- and RP-band data (Fig.~\ref{fig:Light_curves}), which show smaller variations of about 0.1--0.2 magnitudes, suggesting that the star was relatively stable during this time and not undergoing significant color changes. The \textit{Gaia} BP - RP color index remains essentially constant during the time of observation, with a mean of 5.34 mag and a standard deviation of 0.09. This value corresponds to a very reddened object, supporting the high $A_{\rm V}$ reported for the star. The \textit{r}-band and \textit{i}-band light curves from \citet{Hackstein2015AN....336..590H} exhibit larger variations of approximately 0.5--0.6 magnitudes. This order of variability has also been observed in other Galactic LBVs, such as AG Car, where short S Dor (S-SD) cycles are superimposed on the long S Dor (L-SD) cycle \citep{2001A&A...366..508V}. This reinforces the presence of photometric variability in WRAY~16-232, and its similarities to AG Car further support its classification as an LBV.

The bottom panel of the Fig.~\ref{fig:Light_curves} shows a combined light curve including \textit{r}-band, \textit{Gaia} G-band, and \textit{ASAS-SN} V-band data. To enable a visual comparison of variability across different bands, the \textit{ASAS-SN} data have been offset by 3.7 magnitudes and the \textit{r}-band data by 2.2 magnitudes, aligning them with the \textit{Gaia} photometry. This allows for a cohesive plot spanning from 2011 to 2018. A clear similarity in variability trends is observed in the \textit{Gaia} G-band and \textit{ASAS-SN} V-band data, particularly between January 2016 and July 2017, where both light curves show a notable increase in brightness.

We were unable to find long-term photometric observations of WRAY~16-232 that would capture a full L-SD phase, which limits our ability to definitively confirm its status as a LBV. We acknowledge that the variability timescales observed in our analysis are insufficient for a conclusive classification. Nonetheless, the presence of S-SD phase–like variability, coupled with the spectral characteristics discussed in Section~\ref{sec: optical}, provides strong supporting evidence for its classification as an LBV. Regular monitoring of the spectroscopic variability coupled with photometry will help understand the evolution of the star and track possible beginning of S Dor phase.


\subsection{Short Term H$\alpha$ Variations}

\begin{figure}
    \includegraphics[width=1\columnwidth]{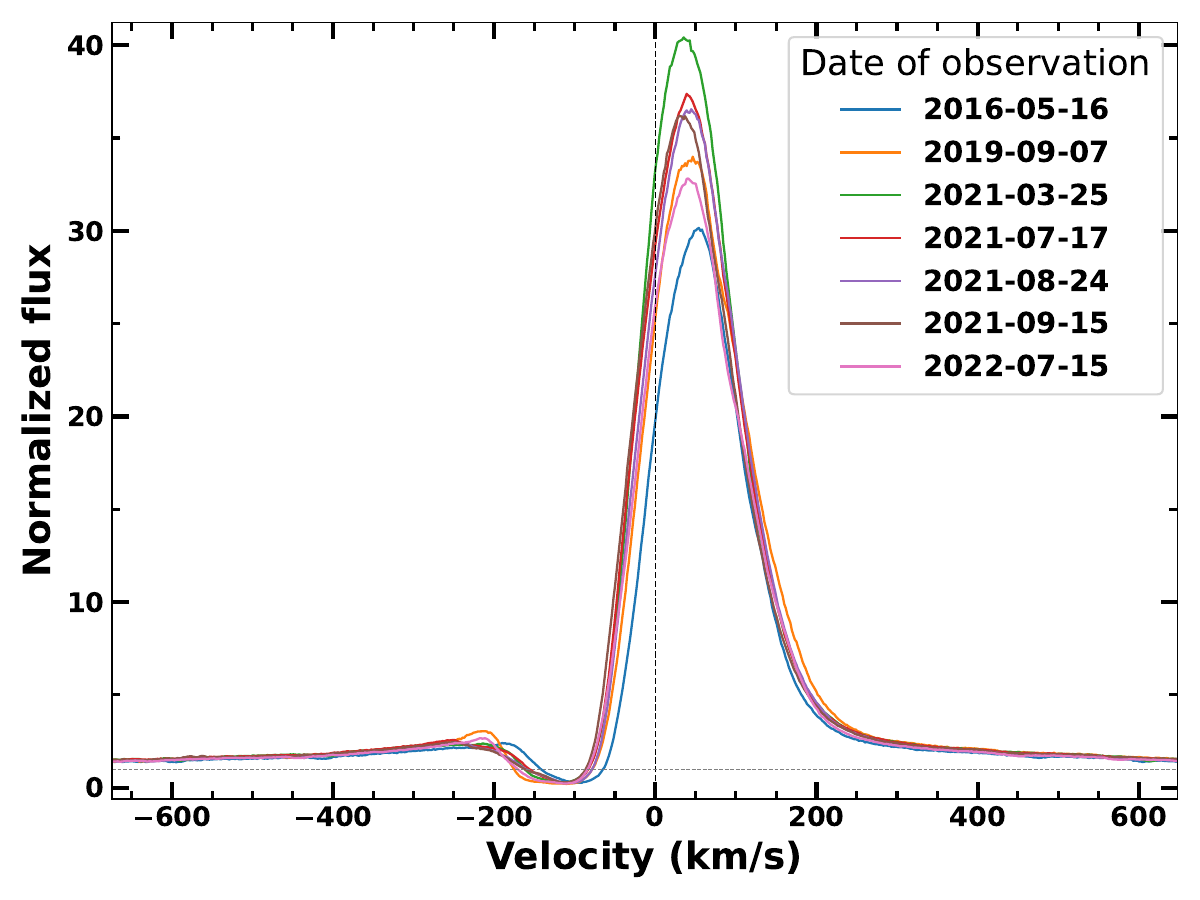}
    \caption{Multi-epoch SALT spectra of WRAY~16-232 plotted on the velocity scale relative to the rest wavelength of H$\alpha$ (6562.8 \AA\, dashed line). Each epoch is shown in a different colour for clarity, and short-term variability in the H$\alpha$ line is clearly evident over the course of the observations.}
    \label{fig:h_alpha_multiepoch}
\end{figure}

We analyzed at the variation of H$\alpha$ in the seven epochs of SALT observation. The spectra reveal short-term variability in the H$\alpha$ line profile (Fig.~\ref{fig:h_alpha_multiepoch}). The morphological changes are clearly evident in the emission peak intensities and in the shape of the wings, suggesting fluctuations on timescales of months to years.

Line velocities, $v_{\infty}$, and EW values are measured for all seven epochs and are summarised in Table~\ref{tab:halpha_measurements}. The H$\alpha$ line velocities remain nearly constant across epochs, with a variability of only $\sim$4\%. In contrast, both the EW (variability $\sim$22\%) and $v_{\infty}$ (variability $\sim$17\%) show significant changes. The terminal velocities, display a clear grouping: in 2021 they cluster around $\sim$280~km~s$^{-1}$, while in 2016, 2019, and 2022 they remain closer to $\sim$250~km~s$^{-1}$. Such shifts in $v_{\infty}$ can be linked to variations in the stellar mass-loss rate. In LBVs, it is well established that visual maximum phases, associated with enhanced mass loss, show reduced terminal velocities, whereas in visual minimum they resemble those of normal supergiants \citep{Groh_2009}. The observed differences therefore suggest higher mass-loss episodes during the 2016, 2019, and 2022 epochs compared to 2021.

\begin{table}
\centering
\caption{Summary of H$\alpha$ measurements at different epochs. Line velocities are reported with $3\sigma$ errors and EW values are given with $1\sigma$ standard deviations.}
\label{tab:halpha_measurements}
\begin{tabular}{lccc}
\hline
Epoch & Line velocity & $v_{\infty}$ & EW \\
      & (km s$^{-1}$) & (km s$^{-1}$) & (\AA) \\
\hline
2016-05-16 &  $150.00 \pm 1.11$ & $244$ & $-27.79 \pm 1.63$ \\
2019-09-07 & $151.20 \pm 1.31$ & $253$ & $-30.01 \pm 2.08$ \\
2021-03-25 & $149.04 \pm 0.97$ & $249$ & $-34.67 \pm 1.78$ \\
2021-07-17 & $151.29 \pm 1.01$ & $289$ & $-31.54 \pm 1.74$ \\
2021-08-24 & $149.26 \pm 1.09$ & $283$ & $-30.95 \pm 1.73$ \\
2021-09-15 & $145.31 \pm 1.02$ & $279$ & $-30.88 \pm 1.44$ \\
2022-07-22 & $151.22 \pm 1.24$ & $255$ & $-29.22 \pm 0.98$ \\
\hline
\end{tabular}
\end{table}

\section{Discussion}\label{sec:discussion}

\subsection{WRAY~16-232: A Promising LBV Candidate} \label{sec: LBV_candidate}

The Spitzer IRAC 4 image (Fig.~\ref{fig:irs_image}) reveals a complex arc-like morphology in the nebula surrounding WRAY~16-232, likely formed by the interaction between faster winds from recent mass-loss events and slower, previously ejected material. The mid-IR Spitzer spectra show forbidden lines of [Fe~\textsc{ii}], [Ne~\textsc{iii}] and [S~\textsc{iii}], closely resembling those of other LBVs known to harbor dust in their environments, such as HR Car \citep{2009ApJ...694..697U} and HD 168625 \citep{Umana2010ApJ...718.1036U}. Additionally, the presence of P Cygni profiles in H~{\sc i}, He~{\sc i}, and Fe~{\sc ii} lines in the optical spectrum, along with a nitrogen-rich nebula indicated by numerous N~{\sc i} lines, further supports its classification as a LBV candidate.

\subsection{Implications of Fullerene Detection} \label{implication_of_fullerene}

A detailed examination of the central, north and south spectral regions, as shown in Fig.~\ref{fig:irs_spectra}, reveals that the central region is devoid of PAHs and C\textsubscript{60}.  The illuminated regions in the IR images appear denser and brighter than their surroundings. It is possible that these clumpy regions are partially shielded from intense UV radiation, giving ideal conditions of complex molecules to form, while still receiving enough exposure to facilitate their excitation. This distinct environment may have contributed to the formation of (i) PAHs through a bottom-up process facilitated by hydrogen-rich conditions, and subsequently (ii) C$_{60}$ through shock-induced top-down fragmentation of larger PAHs observed in the ejecta during the LBV phase (see Section~\ref{sec:c60_survival} and Section~\ref{sec:timescale of C_60 formation} for further discussion).

\begin{figure}
    \includegraphics[width=\columnwidth]{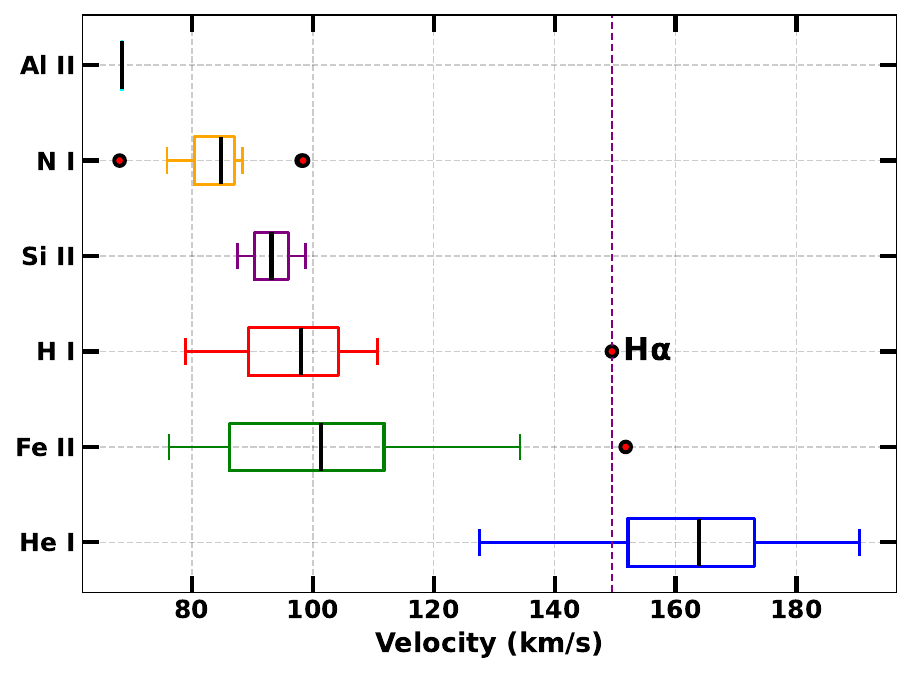}
    \caption{Box plot of the species identified in the optical spectrum of WRAY~16-232. Outliers for each species are indicated by red circles. The black line within each box represents the median velocity, below which 50\% of the transitions for that species are found.}
    \label{fig:vel_box_plot}
\end{figure}

A potential pathway for fullerene formation is suggested in the theoretical work of \citet{Micelotta2010A&A...510A..36M,2010A&A...510A..37M}, which proposes that PAHs could transform into C$_{60}$ through shock-driven processes. Their studies indicate that shock velocities of 75--150\,km\,s$^{-1}$ can significantly modify the structure of PAHs, a velocity range comparable to the slow winds of LBVs. Fig.~\ref{fig:vel_box_plot} presents a velocity box plot of the various species identified in the optical spectrum of WRAY~16-232. Transitions associated with nitrogen, silicon, hydrogen (Paschen series), and iron predominantly fall within the low-velocity range of 75--100\,km\,s$^{-1}$, whereas a higher-velocity range of 120--170\,km\,s$^{-1}$ is associated with He~\textsc{i} and H$\alpha$. Shock regions and shielded dense regions can be created by wind--wind interactions in the shells of LBVs \citep{Weis2020Galax...8...20W}. This suggests that, in the stellar envelope of WRAY~16-232, there may be regions where the velocity conditions facilitate the processing of PAHs into C$_{60}$.

To our knowledge, the theoretical work by \citet{Micelotta2010A&A...510A..36M,2010A&A...510A..37M} has never been tested observationally. WRAY~16-232 and its shell may provide a testing ground to observationally verify fullerene formation through shocks. Furthermore, this could indicate that such PAH-to-fullerene transformation occurs specifically in more LBV-type environments. Here, we explore the physical and temporal constrains of this possible process.

\subsubsection{Physical Constraints on Fullerene Survival}
\label{sec:c60_survival}

Firstly, we estimate the UV radiation field strength ($G_0$) and compare it to known survival thresholds for fullerenes. Fullerenes like C$_{60}$ are resilient in moderate UV environments but undergo photodissociation in intense fields, typically surviving in $G_0 < 10^4$--$10^5$ Habing units (\citealp{Habing1968BAN....19..421H}; \citealp{Berndoi:10.1073/pnas.1114207108}; \citealp{2015A&A...577A.133B}). In PN and RNe where C$_{60}$ is commonly detected, $G_0$ ranges from $10^2$ to $10^5$ Habing units, with higher values requiring additional shielding from gas or dust to prevent destruction \citep{Sellgren_2010}. $G_0$ values in LBVs get significantly reduced with distance and attenuation by dust clumps or shells.

We calculate $G_0$ at the observed shell radius ($r_{\rm shell} \approx 0.7$ pc). For LBV candidates like WRAY~16-232, which exhibit spectral features consistent with a quiescent phase (e.g., T$_{\rm eff} \sim 12,000$--$30,000$ K) from the absence of higher ionization lines and presence of Fe {\sc ii} $L_{\rm bol}$ typically ranges from $10^5$ to $10^6$ L$_{\odot}$ \citep{1994PASP..106.1025H}. This puts our estimates of $G_0$ approximately in the range of  $10^3$--$10^4$ Habing units, which fall within the survival regime for C$_{60}$, particularly in the outer shell where UV dilution is significant. However, near the central star, $G_0$ could exceed $10^5$ Habing units, potentially leading to rapid photodissociation. It matches with the central region spectra which does not show Carbon compounds. Survival thus requires attenuation, consistent with the high extinction (A$_V \approx 12$ mag) and clumpy morphology observed in the IRAC 8 $\mu$m image (Fig.~\ref{fig:irs_image}), which suggests the presence of dense regions ($n_{\rm H} > 10^4$ cm$^{-3}$) providing shielding. 

The presence of the 16.4 $\mu$m PAH feature in the envelope spectra (Fig.~\ref{fig:irs_spectra_contin}) further informs the chemical environment and formation pathway. This band is attributed to large PAHs with $N_{\rm C} \gtrsim 50$--$130$ \citep{Boersma2010A&A...511A..32B}, indicating a carbon- and hydrogen-rich ejecta typical of LBVs undergoing CNO-processed mass loss but retaining sufficient H for PAH formation \citep{Vink2018A&A...615A.119V}. In such environments, C$_{60}$ likely forms via a ``top-down'' process, where these large PAHs are dissociated into smaller fragments (e.g., graphenes or C$_n$ chains) by UV photons or shocks, subsequently isomerizing into cage structures \citep{Bernard-Salas_2012,Omont2021}. This hydrogen rich ejecta makes it less efficient to form "bottom-up" formation of C\textsubscript{60}. The estimated $G_0 \sim 10^3$--$10^4$ Habing units is sufficient for photodissociation of PAHs (threshold $G_0 > 10^3$ Habing units; \citealp{Micelotta2010A&A...510A..36M}) without complete destruction, supporting this mechanism. Alternatively, wind-driven shocks (velocities 75--170 km s$^{-1}$) could enhance dissociation, as modeled by \citet{Micelotta2010A&A...510A..36M}, where $v_{\rm shock} \sim 100$ km s$^{-1}$ yields destruction efficiencies of $\eta \sim 0.1$--$1\%$ per event, allowing gradual processing over multiple episodes.

\subsubsection{Timescales for Top-Down C$_{60}$ Formation} \label{sec:timescale of C_60 formation}
We found that the photodissociation and shock formations are possible in the environment of WRAY 16- 232. To evaluate the feasibility of these top-down C$_{60}$ formation, we estimate the timescales for two competing mechanisms, UV photodissociation and, shock/wind processing and compare them to the LBV phase duration ($\sim 10^4$--$10^5$ years; \citealp{1991IAUS..143..485H}). We estimate the shell expansion timescale to be $ \approx 7,000$ years, considering wind velocity of 100 kms$^{-1}$ with a shell radius of 0.7 pc.

In RNe like NGC 7023, top-down formation occurs via UV photodissociation of large PAHs ($N_{\rm C} \gtrsim 50$--$130$), involving sequential H-loss and C$_2$ ejection. This requires multiple photon absorptions to overcome high activation energies ($\sim 10$--$15$ eV per C$_2$ loss). Models indicate a timescale of $\sim 10^5$ years for complete conversion (e.g., C$_{66}$H$_{20}$ to C$_{60}$) in regions with $G_0 \sim 10^3$--$10^4$ Habing units \citep{Berne2012PNAS..109..401B}, which exceeds the LBV phase duration. The mechanism of photodissociaton is too slow to produce observable amount of C$_{60}$ within the LBV phase.

On the contrary, shock processing in LBVs, driven by wind-wind interaction provides a faster pathway. These shocks recur on S Dor cycles (10 - 20 years), within total LBV timescales of $\sim 10^3$--$10^5$ years. The destruction efficiency of PAHs in shocks is $\sim 20$--$40\%$ C-loss per event for 75--100 km s$^{-1}$ shocks \citep{Micelotta2010A&A...510A..36M}, modeled via nuclear stopping.

For WRAY~16-232, shocks fragment large PAHs (evidenced by the 16.4 $\mu$m band) into C$_{60}$-like structures, with lab analogs showing rapid fullerene production from HACs \citep{García-Hernández_2010,Bernard-Salas_2012}. Cumulative processing over numerous shocks could yield observable C$_{60}$ in $<10^4$ years, fitting the LBV phase and shell expansion. Thus, while UV photodissociation is too slow, shock/wind processing aligns with LBV evolution, producing C$_{60}$ efficiently in the clumpy, carbon-rich shell of WRAY~16-232. 

Our $G_0$ estimates predict that C$_{60}$ is favored in the outer, shielded shell regions, whereas it is destructed in regions close to the star. This supports shock formation from wind-wind interactions, as evidenced by the arc-like structures in the shell (Fig.~\ref{fig:irs_image}). 

To test the wind interaction model of \citep{Micelotta2010A&A...510A..36M}, high-resolution integral field spectroscopy and imaging in the near and mid infrared wavelengths are essential. This presents a strong case for obtaining high-resolution James Webb Space Telescope (JWST; \citealp{Gardner_2006}) observations of WRAY~16-232, which would allow us to directly investigate the proposed C$_{60}$ formation mechanisms, resolve $G_0$ gradients in the shell and could potentially target shock tracers (e.g,: H\textsubscript{2}S(0)-S(3) lines).

This is the first ever detection of C$_{60}$ in a LBV environment and may have huge implications on the formation of the species. Fullerenes may be forming in the ejecta during episodic mass-loss events through shocks or through bottom-up processes specific to the LBVs or other massive stellar environments. As massive stars with relatively short lifetimes, LBVs, especially WRAY~16-232, provide a unique opportunity to study the timescales and mechanisms of C$_{60}$ formation. 


\section{End Remarks and Future Work}\label{sec:conclusion}

The present study of WRAY~16-232 supports its previously proposed classification as a strong LBV candidate while also revealing that its circumstellar environment exhibits unique chemical signatures. Our analysis, based on Spitzer MIR spectra, SALT optical HRS spectra, and photometric data from Gaia, ASAS-SN and Bochum GDS has highlighted the existence of complex mass-loss dynamics and dust formation processes. 

The results provide new insights into the interplay between stellar winds, dust formation, and molecular chemistry in LBV shells. Moreover, with the detection of C$_{60}$ molecule in the environment of WRAY~16-232, we find observational  evidence for formation pathway for C$_{60}$ in LBV environments through shock driven process. This discovery has implication to carbon dust formation in the outer shells of massive stars. 

Further investigations involving long term multi-epoch high resolution spectroscopy and high cadence photometry would help in investigating a complete S-Dor cycle of WRAY~16-232. Given its carbonaceous molecular complexity and expanding stellar winds, WRAY~16-232 emerges as a ideal candidate for future high resolution observations using JWST.

\section*{Acknowledgements}
The authors thank the referee for the meticulous review. S.A P thanks the BGS, Director and Dean Indian Institute of Astrophysics (IIA), for providing the opportunity to work under the Visiting Student Program (VSP) at IIA. BM acknowledge the financial support  by the SERB project (CRG/2023/005271). Some of the observations reported in this  paper were obtained with the Southern African Large Telescope (SALT) under program [2016-1-SCI-012 - PI:Alexei Kniazev]. This work has made use of data from the European Space Agency (ESA) mission
{\it Gaia} (\url{https://www.cosmos.esa.int/gaia}), processed by the {\it Gaia} Data Processing and Analysis Consortium (DPAC, \url{https://www.cosmos.esa.int/web/gaia/dpac/consortium}). Funding for the DPAC has been provided by national institutions, in particular the institutions participating in the {\it Gaia} Multilateral Agreement. Also, ChatGPT (OpenAI 2023) is used for assistance in correcting typos and grammar, though full responsibility for the manuscript’s content remains our own.

\section*{Data Availability}

The Spitzer IRS data from the CASSIS can be accessed at \url{https://cassis.sirtf.com/} and CASSISJuice\footnote{https://gitlab.com/cassisjuice}. The complete results produced by this study will be made available to interested parties upon request to the corresponding author


\bibliographystyle{mnras}
\bibliography{sources}


\newpage
\clearpage 
\onecolumn
\appendix


\clearpage
\newpage 
\section{Detected lines} \label{line_velocity}
We present the detected spectral lines with their observed and laboratory wavelengths, noting whether they show P Cygni profiles in \autoref{tab:append}. Line identifications follow \citet{NIST_ASD}. For P Cygni lines that are clearly resolved and unblended, we also provide accurate terminal velocity measurements. Velocity uncertainties were estimated from the photon-noise limit, following \citet{Bouchy2001A&A...374..733B}. 
For a given wavelength region, we measured the local continuum signal-to-noise ratio (S/N) and used it to compute the velocity error as $\sigma_v = \frac{c/R}{\mathrm{S/N}}$, where $c$ is the speed of light and $R$ is the resolving power of SALT HRS ($\approx 14{,}000$ in low-resolution mode). Each measured S/N value was then applied to all spectral lines within the corresponding wavelength interval. This procedure yields the 
minimum statistical error expected for our spectra. We report three times this value ($3\sigma_v$) as the velocity error.

\clearpage
\begin{center}
\renewcommand{\arraystretch}{1.3} 
\setlength{\tabcolsep}{13pt} 

\begin{longtable}{c|c|c|c|c}
\caption{Identified spectral lines in the optical spectrum of WRAY~16-232 within the wavelength range of 5500--8800 \AA. Column 1 indicates the species, Column 2 shows the observed wavelength in the spectrum, and Column 3 provides the corresponding laboratory wavelength (in air) from \citet{NIST_ASD}. Column 4 specifies whether the line exhibits a P Cygni profile, and Column 5 presents the velocity measured for the clearly distinguishable P Cygni lines.} \label{tab:append} \\
\hline
\hline
Species & Observed Wavelength & Laboratory Wavelength & P Cygni Profile & Velocity \\
        & (\AA)              & (\AA, Air)           &                 & (km s$^{-1}$) \\
\hline
\endfirsthead

\hline
\hline
Species & Observed Wavelength & Laboratory Wavelength & P Cygni Profile & Velocity \\
        & (\AA)              & (\AA, Air)           &                 & (km s$^{-1}$) \\
\hline
\endhead

\hline
\endfoot

\hline
\endlastfoot

Fe II & 5535.00 & 5534.89 & Yes & \\
{[Fe II]} & 5747.48 & 5746.97 & No & \\
{[N II]} & 5754.48 & 5754.59 & No & \\
Fe II & 5813.58 & 5813.66 & Yes & 92 $\pm$ 4.95 \\
Fe II & 5835.28 & 5835.45 & Yes & \\
He I & 5876.66 & 5877.25 & Yes & 190 $\pm$ 4.95 \\
Fe II & 5952.70 & 5952.51 & Yes & 106 $\pm$ 4.95 \\
Si II & 5957.90 & 5957.56 & No & \\
Si II & 5979.04 & 5978.93 & No & \\
Fe II & 5991.26 & 5991.37 & Yes & \\
{[Fe II]} & 6045.06 & 6044.10 & No & \\ 
Fe II & 6084.24 & 6084.10 & Yes & \\
Fe II & 6103.22 & 6103.90 & Yes & 98 $\pm$ 4.95 \\
Fe II & 6149.06 & 6149.23 & Yes & 152 $\pm$ 4.95 \\
Fe II & 6179.46 & 6179.37 & Yes & 104 $\pm$ 4.95 \\
Fe II & 6238.95 & 6238.37 & Yes & 127 $\pm$ 1.62 \\
Fe II & 6247.60 & 6247.55 & Yes & 111 $\pm$ 1.62 \\
Fe II & 6305.21 & 6305.29 & Yes & 99 $\pm$ 1.62 \\
Fe II & 6317.93 & 6318.01 & No & \\
Fe II & 6331.92 & 6331.95 & Yes & 82 $\pm$ 1.62 \\
Si II & 6347.29 & 6347.10 & Yes & 99 $\pm$ 1.62 \\
Si II & 6371.60 & 6371.36 & Yes & 88 $\pm$ 1.62 \\
Fe II & 6383.82 & 6383.72 & Yes & \\
Fe II & 6385.78 & 6385.45 & - & \\
Fe II & 6416.92 & 6416.93 & Yes & 89 $\pm$ 1.62 \\
Fe II & 6432.66 & 6432.67 & Yes & 112 $\pm$ 1.62 \\
Fe II & 6456.39 & 6456.38 & Yes & 134 $\pm$ 1.62 \\
N II & 6482.28 & 6482.05 & Yes & 87 $\pm$ 1.62 \\
Fe II & 6491.26 & 6491.25 & - & \\
Fe II & 6492.79 & 6493.04 & - & \\
Fe II & 6515.87 & 6516.07 & Yes & \\
Fe II & 6516.85 & 6517.02 & Yes & \\
H$\alpha$ & 6563.45 & 6562.79 & Yes & 150 $\pm$ 1.11 \\
C II & 6582.96 & 6582.87 & No & \\
{[N II]} & 6582.96 & 6583.45 & No & \\
Fe II & 6626.94 & 6627.23 & Yes & 81 $\pm$ 1.11 \\
He I & 6678.93 & 6679.99 & Yes & 160 $\pm$ 1.05 \\
{[S II]} & 6717.19 & 6716.44 & No & \\
Al II & 6823.56 & 6823.48 & No & \\
N II & 6966.86 & 6966.81 & Yes & 98 $\pm$ 0.48 \\
Al II & 7042.29 & 7042.06 & Yes & 69 $\pm$ 0.48 \\
Al II & 7056.66 & 7056.60 & Yes & 68 $\pm$ 0.48 \\
He I & 7065.77 & 7065.19 & Yes & 167 $\pm$ 0.48 \\
Fe II & 7134.77 & 7134.54 & Yes & 85 $\pm$ 0.48 \\
{[Fe II]} & 7155.23 & 7155.17 & No & \\
Fe II & 7222.21 & 7222.39 & Yes & 93 $\pm$ 0.42 \\
Fe II & 7320.49 & 7320.70 & Yes & 80 $\pm$ 0.42 \\
N I & 7423.91 & 7423.64 & Yes & 84 $\pm$ 0.42 \\
N I & 7442.38 & 7442.29 & Yes & 80 $\pm$ 0.42 \\
Fe II & 7449.42 & 7449.34 & Yes & 98 $\pm$ 0.42 \\
Fe II & 7462.39 & 7462.41 & Yes & 109 $\pm$ 0.42 \\
N I & 7468.89 & 7468.31 & Yes & 98 $\pm$ 0.42 \\
Fe II & 7479.58 & 7479.70 & Yes & 76 $\pm$ 0.42 \\
Fe II & 7494.94 & 7494.69 & No & \\
Fe II & 7495.69 & 7495.62 & No & \\
Fe II & 7506.62 & 7506.54 & - & \\
Fe II & 7513.33 & 7513.17 & No & \\
Fe II & 7515.15 & 7515.10 & No & \\
Fe II & 7533.34 & 7532.50 & Yes & 109 $\pm$ 0.42 \\
Fe II & 7711.60 & 7711.43 & Yes & 113 $\pm$ 0.45 \\
Fe II & 7731.68 & 7731.68 & No & \\
Fe II & 7755.63 & 7756.00 & No & \\
O I & 7772.11 & 7771.94 & - & \\
O I & 7775.67 & 7775.39 & Yes & \\
Fe II & 7789.13 & 7789.27 & No & \\
Fe II & 7801.27 & 7801.19 & No & \\
Fe II & 7818.16 & 7818.32 & No & \\
Si II & 7848.94 & 7849.72 & No & \\
Si II & 7849.52 & 7848.80 & No & \\
Fe II & 7866.54 & 7866.55 & No & \\
Mg II & 7877.06 & 7877.05 & No & \\
Mg II & 7896.49 & 7896.36 & No & \\
Fe II & 7917.86 & 7917.79 & No & \\
Fe II & 7975.85 & 7975.91 & No & \\
Fe II & 7981.77 & 7981.91 & No & \\
H I (P35) & 8264.57 & 8264.28 & Yes & \\
H I (P29) & 8292.61 & 8292.30 & Yes & \\
H I (P28) & 8298.41 & 8298.83 & Yes & \\
H I (P27) & 8306.15 & 8306.10 & Yes & \\
H I (P26) & 8314.60 & 8314.26 & Yes & \\
H I (P25) & 8324.34 & 8323.42 & Yes & 105 $\pm$ 0.24 \\
H I (P24) & 8334.07 & 8333.78 & Yes & \\
H I (P23) & 8345.75 & 8345.54 & Yes & 88 $\pm$ 0.24 \\
H I (P22) & 8359.13 & 8359.00 & Yes & 79 $\pm$ 0.24 \\
H I (P21) & 8375.07 & 8374.48 & Yes & 102 $\pm$ 0.24 \\
H I (P20) & 8392.76 & 8392.40 & Yes & 100 $\pm$ 0.24 \\
H I (P19) & 8413.68 & 8413.32 & Yes & 92 $\pm$ 0.24 \\
H I (P18) & 8438.11 & 8437.95 & Yes & \\
O I & 8446.60 & 8446.36 & Yes & \\
Fe II & 8450.91 & 8451.01 & Yes & \\
H I (P17) & 8468.47 & 8467.26 & Yes & \\
Ca II & 8497.92 & 8498.02 & Yes & \\
H I (P16) & 8502.72 & 8502.49 & Yes & \\
Ca II & 8542.19 & 8542.09 & Yes & \\
H I (P15) & 8545.43 & 8545.38 & Yes & \\
N I & 8567.79 & 8567.74 & Yes & 83 $\pm$ 0.24 \\
He I & 8582.95 & 8582.62 & Yes & 128 $\pm$ 0.24 \\
H I (P14) & 8598.77 & 8598.39 & Yes & 96 $\pm$ 0.24 \\
N I & 8629.30 & 8629.24 & Yes & \\
Ca II & 8662.18 & 8662.14 & Yes & \\
H I (P13) & 8665.16 & 8665.02 & Yes & \\
N I & 8680.24 & 8680.28 & Yes & \\
N I & 8683.14 & 8683.40 & Yes & \\
N I & 8686.33 & 8686.15 & Yes & \\
N I & 8703.48 & 8703.25 & Yes & 87 $\pm$ 0.24 \\
N I & 8712.01 & 8711.70 & Yes & 87 $\pm$ 0.24 \\
N I & 8719.13 & 8718.83 & Yes & 88 $\pm$ 0.24 \\
N I & 8729.03 & 8728.89 & Yes & 81 $\pm$ 0.24 \\
H I (P12) & 8751.44 & 8750.46 & Yes & 111 $\pm$ 0.24 \\
\end{longtable}

\end{center}

\clearpage
\bsp	
\label{lastpage}
\end{document}